\documentclass[dvipsnames,twocolumn]{aastex61}

\usepackage{amsmath}
\usepackage{myshorthand}
\usepackage[caption=false]{subfig}
\renewcommand{\revonecolor}{} % \color{NavyBlue}
\renewcommand{\revtwocolor}{} % ..some purple

\begin{document}
\title{Deconvolving the HD 81809 binary: rotational and activity evidence for a subgiant with a Sun-like cycle}
\author[0000-0002-4996-0753]{Ricky Egeland}
\affiliation{High Altitude Observatory, National Center for Atmospheric Research‡, PO Box 3000, Boulder, CO 80307-3000, USA}
\correspondingauthor{Ricky Egeland}
\email{egeland@ucar.edu}

\begin{abstract}
\revone{
HD 81809 has one of the highest quality activity cycles from the
sample of stars synoptically observed in the Mount Wilson Observatory
HK Project.  However, this object is in fact a binary system, raising
the question as to which of the components is responsible for the
observed cyclic activity and what are the properties of that active
component.  The \emph{Hipparcos} spacecraft obtained resolved
two-color photometry for this system that indicates that both
components are near the solar temperature.  Combined with the precise
\emph{Gaia} parallax and empirical bolometric corrections we derive
component luminosities of $L_A = 5.8 \pm 0.3 \, \Lnom$ and $L_B =
1.025 \pm 0.055 \, \Lnom$, and radii $R_A = 2.42 \pm 0.08 \, \Rnom$
and $R_B = 1.04 \pm 0.04 \, \Rnom$, confirming that the primary
component is a subgiant.  We perform an independent estimate of the
rotation period of the A component based on $\vsini$ and find that it
agrees with the $40.2$ d period previously measured from the Ca HK
time series.  We explore plausible scenarios for the deconvolved
$S$-index and find that a cycling A component would have an activity
level within the bounds of ensemble activity-rotation trends, while a
cycling B component likely does not.  Based on the available rotation
and activity evidence, we find the most likely characterization of the
system is a subgiant primary component responsible for the smooth
cyclic behavior in Ca HK with $\log(\RpHK) \sim -4.89$, while the
secondary component has relatively flat activity at $\log(\RpHK) \sim
-5.02$.
}
\end{abstract}


%%%
%%%
%%%
\section{Introduction}

Diligent observations of magnetic activity proxies for ``Sun-like''
stars have enabled astronomers to search for stellar counterparts to
the long-studied solar sunspot cycle.  Olin C. Wilson and colleagues
were the first to commit to such an observation program with the Mount
Wilson Observatory (MWO) HK Project using the HKP-1 spectrophotometer
\citep{Wilson:1968,Wilson:1978}.  The activity proxy of choice was the
emission in the calcium H \& K line cores produced in the lower
chromosphere by magnetic heating processes
\citep{Linsky:1970}.  Wilson's observational records were
extended by the HKP-2 photometer \citep{Vaughan:1978} and after 25
years of observation a variety of activity patterns were discovered in
magnetically active stars \citep{Baliunas:1995}.  Besides the familiar
pattern of the quasi-periodic solar cycle
\citep{Livingston:2007,Egeland:2017}, stars were found with erratic,
non-periodic variability or nearly flat activity records.  The
existence of these varieties of variability raises the question ``What
are the physical conditions conducive to producing a smoothly varying
quasi-periodic magnetic cycle like that of the Sun?''

We can make progress on this question in a statistical sense by
carefully analyzing the fundamental properties of a sample of stars
whose long-term activity is observed.  This is a small sample of at
most a few hundred stars at present, due to the difficulty of
maintaining such long-term observation programs in a social and
scientific environment that prioritises short-term advances.  The MWO
HK Project endured the longest from 1966--2003.  Using 25 years of
data for 111 FGK-type stars, \citep{Baliunas:1995} found that about
half of their sample had statistically significant evidence for a
periodic cycle, about a quarter were erratic variables, and another
quarter had flat (or nearly so) activity.  The cycling class was
further divided into four classes based on their estimated quality:
poor, fair, good, and excellent, with the Sun demonstrating an
``excellent'' cycle.  These classes have a noticeably different
character, with the ``excellent'' and ``good'' classes showing obvious
periodicity in their time series like the Sun, while the ``fair'' and ``poor''
classes are not so obvious and indeed it is questionable whether they
are actually Sun-like cycles at all.  For this reason, subsequent
studies of the relationship between cycle period and stellar
properties such as rotation and convective turnover time typically
use only the high-quality ``excellent'' and ``good'' cycles
(\citet{Saar:1999}, \citet{Brandenburg:1998},
\citet{Bohm-Vitense:2007}, \citet{Brandenburg:2017})

%%  The authors noted a correspondence between cycle
%% length and quality, finding that no short ``cycles'' ($< 7$ years) were in
%% the ``good'' or ``excellent'', and suggested that these short ``cycles''
%% may not persist throughout their 25 year records.  (Note that it is
%% typical in the literature to call all statistically significant peaks
%% in a periodogram analysis ``cycles'', regardless of whether they are
%% persistent.)

\citet{Baliunas:1995} observed that ``K-type stars with low
$\anglemean{S}$ almost all have pronounced cycles.''  But what about
G-type stars, like the Sun?  \citet{Egeland:2018b} conducted a
statistical analysis of the \citep{Baliunas:1995} sample, finding that
only 3 stars in the G-group ($0.58 \leq (B-V) < 0.75$) have
high-quality ``excellent'' or ``good'' cycles, including the Sun.  The
vast majority of the other high-quality cycles (17/21) are from the
K-group ($(B-V) \geq 0.75$).  The other G-group high-quality cyclers
include HD 78366 ($(B-V) = 0.60$) and HD 81809 ($(B-V) = 0.65$).  The
former has two significant cycle periods and therefore does not have
variability qualitatively similar to the Sun, and furthermore is
evolved (luminosity class IV-V; \citet{Gray:2003}).  The latter is the
subject of this work; it has a very Sun-like $\sim$8 yr ``excellent''
class cycle, however its characterization is uncertain because it is
in fact an unresolved binary system with a convolved $(B-V)$ of 0.65.
Because the cycle of the convolved Ca HK signal for HD 81809 is so
pronounced, it is clear that the variability is dominated by only one
component of the binary.  The other component is either flat activity,
or has an amplitude of variability in HK flux that is much smaller
than the cycling component.  In summary, unless the variable component
of HD 81809 turns out to be a G-type star, then \emph{there are no
 other G-type stars in the \citet{Baliunas:1995} sample which have a
Sun-like cycle.}  \citet{Egeland:2018b} concluded that the Rossby
number, the ratio of the rotation period to the convective turnover
timescale, appears to be the critical parameter which determines
whether a star shall have a Sun-like cycle. The relatively large solar
Rossby number is found to be close to that of the K-type stars with
high-quality cycles.

HD 81809 was apparently not known to be a binary by Olin Wilson at the
onset of the observation program in 1966 \citep{Baliunas:1998},
although it has been observed as a visual binary such since at least
1938 \citep{VanDenBos:1938,Baize:1985}.  With only 12 years of data
\citet{Wilson:1978} found HD 81809 to be the only star with
``definitely'' cyclic behavior with a period of about 10 years, and
remarked that the star is ``rather similar to the Sun in spectral type
and [cycle] period''. Visual measurements
by \citet{Baize:1985} found a separation of $\sim$0.4\arcsec{}, later
confirmed with speckle interferometry \citep{McAlister:1990}, and
an orbital period of 35 yr.  Radial
velocity measurements by \citet{Duquennoy:1988} were consistent with
the orbital period of 35 yr and component masses of $1.5 \pm 0.5 M_\Sun$
and $0.8 \pm 0.2 M_\Sun$.  The rotation period measured from the Ca HK
time series was found to be quite long, $\sim$40 d \citep{Noyes:1984},
which led \citet{Baliunas:1995} to suggest that the variations were
due to the lower-mass component with an estimated spectral type of
$\sim$K0.  This was motivated by the activity-rotation study of
\citet{Noyes:1984}, in which such a slow-rotating, moderately active
($\anglemean{S} = 0.172$) star would be peculiar were the higher-mass
component (estimated to be $\sim$G0V) responsible for the variability.
Later works continued to assume the HK emission of HD 81809 was due to
a K-type component with $(B-V) \sim 0.80$
\citep[e.g.][]{Brandenburg:1998,Saar:1999,Bohm-Vitense:2007}.  One
notable divergence from this interpretation comes from
\cite{Favata:2008}, who studied the X-ray emission of HD 81809 and
concluded that only the larger, higher mass component has enough
surface area to produce the observed X-ray fluxes assuming solar-like
active regions.  We will further discuss what we can conclude from the
X-ray observations of HD 81809 in Section \ref{sec:discussion}.

The purpose of this work is to use the latest available observations 
to characterize the components of HD 81809, and from there infer which
of the components is most likely responsible for the Ca HK variations
resulting in an $\sim$8 yr cycle and $\sim$40 day rotation.  In
Section \ref{sec:obs} we present the \emph{resolved} two-color photometry from the Tycho
instrument on \emph{Hipparcos}, and how these data along with the
observed orbital properties update our understanding of the
components.  Next, in Section
\ref{sec:rot} we develop an argument based on rotation that suggests
the A component is responsible for the Ca HK observations.  In Section
\ref{sec:deconv} we attempt to deconvolve the observed $S$-index based
on the resolved $(B-V)$.  \revone{Finally, we summarize our
  conclusions in Section \ref{sec:conclusions} and end 
with a discussion of the implications of our new
characterization of HD 81809 in Section \ref{sec:discussion}.}

%%%
%%%
%%%
\section{Observations and Characterization of the HD 81809 System}
\label{sec:obs}

%%%
\subsection{Ca \II{} HK Activity Cycle and Rotation}

\begin{figure*}
\centering
    \includegraphics[width=\textwidth]{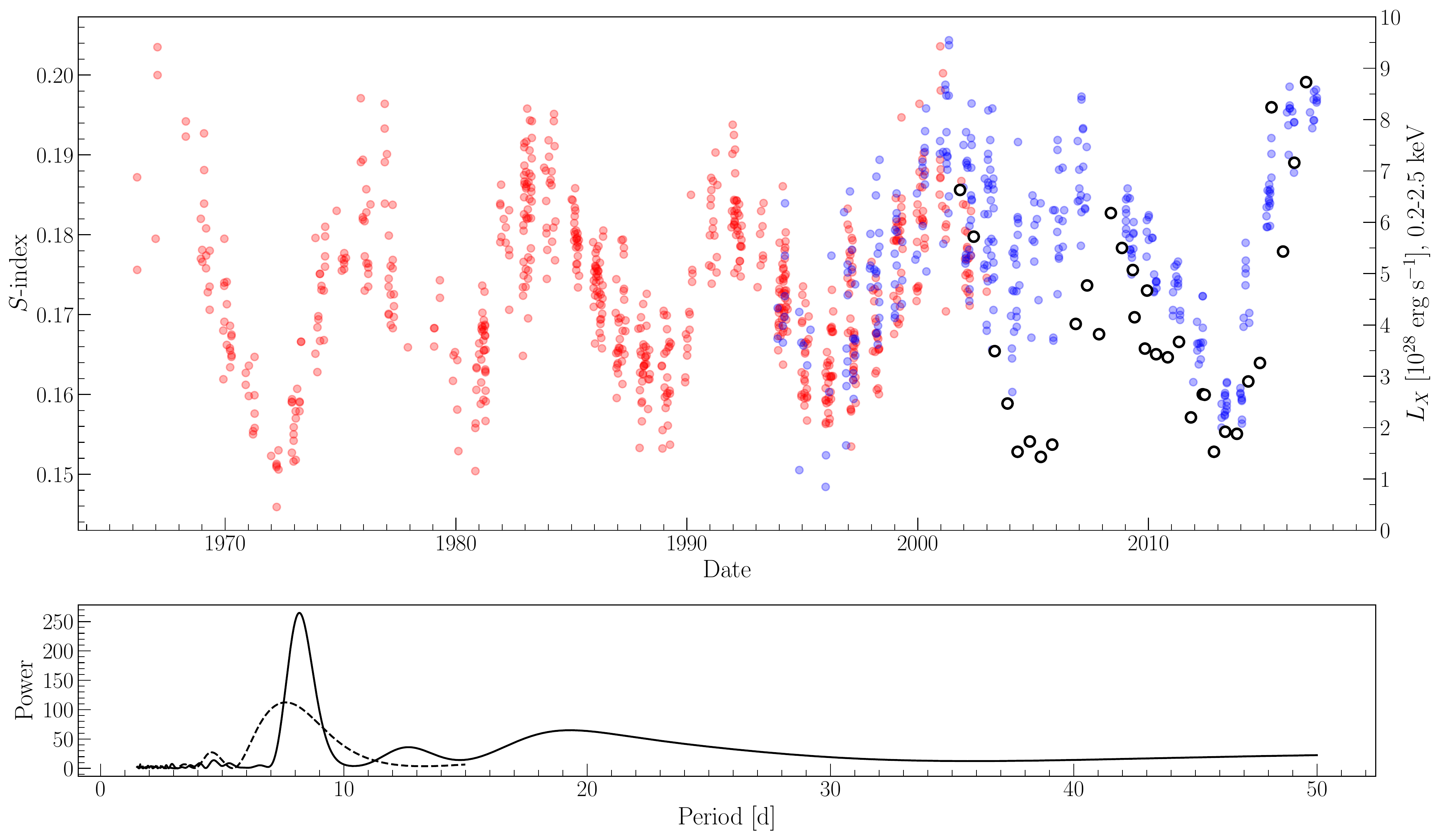}
    \caption{\revtwo{\emph{Top:}} Composite $S$-index time series for HD 81809, with MWO
      observations in red and SSS in blue.  XMM-Newton X-ray
      observations are overplotted as open circles with the scale on
      the right axis.  \revtwo{\emph{Bottom:} Lomb-Scargle periodogram of
      the composite $S$-index time series (solid line) and X-ray
      observations (dashed line), the latter multiplied by 10 for
      better visibility.}}
    \label{fig:timeseries}
\end{figure*}

Figure \ref{fig:timeseries} shows the 51-year composite $S$-index time
series for the HD 81809 system, with observations from the MWO HK
Project in red and those from the Lowell Observatory Solar-Stellar
Spectrograph (SSS) in blue.  SSS observations are calibrated to the
Mount Wilson scale using a linear regression for near-coincident
observations from an ensemble of 26 solar-analog stars and an
additional scaling factor to eliminate remaining discontinuity
\citep{Egeland:2017:thesis}.  $S$ is defined as:

\begin{equation}\label{eq:S}
  S = \alpha \frac{N_H + N_K}{N_R + N_V}
\end{equation}

\noindent where $N_H$ and $N_K$ are the counts in 1.09 \AA{}
triangular bands centered on Ca \II{} H \& K in the HKP-2
spectrophotometer, $N_R$ and $N_V$ are 20 \AA{} reference bandpasses
in the nearby continuum region, and $\alpha$ is a calibration constant
\citep{Vaughan:1978}.  The binary components are not resolved in these
observations, therefore flux in each of the four bands is convolved.
This \revone{long} time series clearly shows five full and two partial
Sun-like cycles.  A Lomb-Scargle periodogram analysis
\citep{Lomb:1976,Horne:1986} \revtwo{of the composite $S$-index data
  finds the cycle period to be $8.17 \pm 0.02$ yr, where we have
  estimated the uncertainty using equation (3) of
  \citet{Baliunas:1995}, and furthermore have found our cycle period
  estimate using an additional 25 years of observations to be
  equivalent to their previous value of $8.17 \pm 0.08$.}  The median
activity is $\median{S} = 0.1751 \pm 0.0003$, where the 1-$\sigma$
uncertainty is estimated using the median absolute deviation scaled on
the assumption of a Gaussian distribution.\footnote{\revtwo{In this
    work we use $\median{x}$ to denote median of a series of values
    $x$, while the mean is denoted $\anglemean{x}$.}}

\cite{Donahue:1993:thesis} \revtwo{and \cite{Donahue:1996}}
conducted a periodogram analysis on seasonally binned data from MWO
available at the time, finding statistically significant periods in 6
of 26 seasons.  The mean rotation period \revone{\citeauthor{Donahue:1993:thesis}} found was 40.2 days with an
rms of 2.3 days, and with seasonal measurements ranging from 37.0 to 43.0
days.  The maximum difference $\Delta P = 6.0$ days can be interpreted
as a sign of differential rotation due to rotational modulations
originating from active regions at varied latitudes
\citep{Donahue:1996}.  \revone{ \citet{Donahue:1993:thesis}
also looked for correlations between seasonal rotation period
measurements and seasonal mean activity that could reveal a solar-like
pattern of slow rotation at times of low activity when sunspots appear
at high latitudes.  No such pattern was found for HD 81809.}
\revtwo{An updated rotation search is planned for HD 81809 and an
  ensemble of MWO and SSS stars in a future publication.  For the present
  work we will rely on the results of \citet{Donahue:1996}.}

Figure \ref{fig:timeseries} also shows the X-ray 0.2--2.5 keV
luminosity timeseries of \citet{Orlando:2017} observed by
\emph{XMM-Newton} (excluding one high $L_X$ observation presumably due
to a flare).  The X-ray observations presently cover one whole and two
half-cycles, and is clearly in phase with the Ca HK cycle.  The median
$L_X$ is $3.52 \times 10^{28}$ erg s$^{-1}$, and the range is from
1.43 to 8.73 $\times 10^{28}$ erg s$^{-1}$.  \citep{Judge:2003}
estimated the solar X-ray flux in the \emph{ROSAT} 0.1--2.4 keV
bandpass from \emph{SNOE-SXP} observations, finding it to vary from
$6.31 \times 10^{26}$ to $7.94 \times 10^{27}$ erg s$^{-1}$ over the
solar cycle with an absolute accuracy of $\pm 50$\%.  We estimate the
median solar luminosity from this to be the midpoint, $4.29 \times
10^{27}$ erg s$^{-1}$.  Neglecting the small offset in the bandpass,
the median X-ray luminosity of HD 81809 is a factor 8.2 larger than
the Sun and the (max $-$ min) amplitude is 10 times larger.  \revtwo{A
  Lomb-Scargle periodogram of the X-ray data finds a cycle period of
  $7.6 \pm 0.2$ yr.  This is notably shorter than the period derived
  from the $S$-index time series, which could be due to the cycle
  covered by the X-ray observations being shorter than the mean.  To
  test this, we computed a Lomb-Scargle periodogram of an $S$-index
  time series truncated to within the time limits of the X-ray
  observations.  In that case, we obtained a cycle period of $7.9 \pm
  0.2$ yr, which is compatible to within the uncertainties with the
  X-ray cycle.}\footnote{\revtwo{\citet{Orlando:2017} reported an
  X-ray cycle period of $7.3 \pm 0.3$ yr by fitting a sinusoid to
  their data.  We were not able to reproduce this result from their
  published data and using the Lomb-Scargle periodogram.  However,
  by including the presumably flaring observation of 2002-11-02 we
  obtained a shorter cycle period of 7.2 yr.}}

%%%
\subsection{Orbit Observations and Component Masses}

\begin{figure}
  \label{fig:orbits}
  \centering
  \subfloat{
    \includegraphics[width=0.8\linewidth]{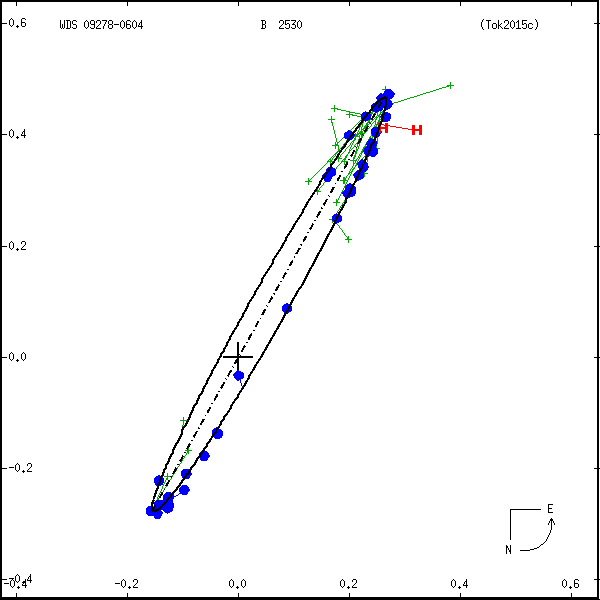}
    \label{fig:astrometric_orbit}
  }
  \\
  \subfloat{
    \includegraphics[width=\linewidth]{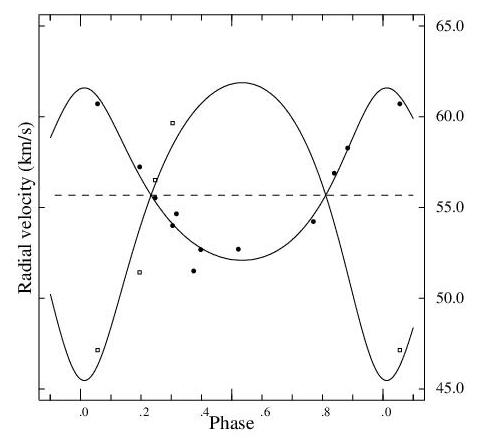}
    \label{fig:RV_orbit}
  }
  \caption{Orbital observations.  \emph{Top}: Orbital solution and astrometry from
    \citet{Tokovinin:2015}.  The position of the primary at the
    coordinate center is marked by a large cross, the line of nodes is
    traced by a dash-dotted line, and the orbital solution a solid
    line. The scale is in arcseconds. Interferometric (solid blue),
    \emph{Hipparcos} (red), and micrometer (green crosses) measures are
    connected to their expected positions on the calculated
    orbit. \emph{Bottom}: Radial velocity observations from $S_{B^9}$
    catalog \citep{Pourbaix:2004}.  Filled circles are the primary and
    open circles are the secondary.  The solid lines show the orbital
    solution, while the dashed line gives the center-of-mass radial
    velocity.}
\end{figure}

\begin{deluxetable}{lcc}
\tablecaption{HD 81809 Orbital and Blended Properties\label{tab:system}}
\tablehead{
  \colhead{Property} & \colhead{Value} & \colhead{Reference}
}
\startdata
parallax, $\pi$           & \revone{$29.63 \pm 0.76$} mas & (1) \\
semimajor axis, $a$       & $0.428 \pm 0.001$ as & (2) \\
                          & \revone{$14.44 \pm 0.37$} AU  &     \\
orbital period, $P_\orb$  & $34.80 \pm 0.06$ yr  & (2) \\
orbital inclination, $i_\orb$ & $85.4 \pm 0.1$ \degrees  & (2) \\
$B$                       & 6.02                 & (3) \\
$V$                       & 5.38                 & (3) \\
$B-V$                     & 0.64                 & (3) \\
$B_T$                     & $6.196 \pm 0.014$    & (4) \\
$V_T$                     & $5.483 \pm 0.009$    & (4) \\
$q = M_A/M_B$             & $0.579 \pm 0.084$    & (5) \\
$M_A + M_B$               & $2.7 \pm 0.9$        & (5) \\
                          & \revone{$2.5 \pm 0.2$}        & (E18) \\
{[Fe/H]}                  & $-0.29$              & (6) \\
\vsini                    & $2.9 \pm 0.3$ km/s   & (7) \\
rotation period, $P_\rot$ & $40.2 \pm 2.3$ d     & (8) \\
cycle period, $P_\cyc$    & \revtwo{$8.17 \pm 0.02$} yr   & (E18) \\
\revtwo{HK} activity, $\median{S}$ & \revtwo{$0.1751 \pm 0.0003$} & (E18) \\
\revtwo{X-ray activity, $\median{L_X}$} & \revtwo{$3.52 \times 10^{28}$ erg s$^{-1}$} & \revtwo{(9, E18)} \\
\enddata
\tablerefs{
(1) Gaia DR2 \citep{Gaia:2018:DR2}
(2) \citet{Tokovinin:2015}
(3)  \citet{Johnson:1966}
(4) \citet{Hog:2000}
(5) \citet{Pourbaix:2000} + $S_{B^9}$ website 
(6) \citet{Holmberg:2009}
(7) \citet{AmmlerVonEiff:2012}
(8) \citet{Donahue:1993:thesis} and \citet{Donahue:1996}
(9) \citet{Orlando:2017}
(E18) This work.
}
\end{deluxetable}

\begin{deluxetable}{lccc}
\tablecaption{HD 81809 Component Properties\label{tab:components}}
\tablehead{
  \colhead{Property} & \colhead{A Value} & \colhead{B Value} & \colhead{Reference}
}
\startdata
$Hp$              & $5.746 \pm 0.003$ & $7.25 \pm 0.01$  & (1)   \\
$B_T$             & $6.34 \pm 0.01$ & $8.25 \pm 0.02$    & (2)   \\
                  & \multicolumn{2}{c}{$6.17 \pm 0.03$}  &       \\
$V_T$             & $5.63 \pm 0.01$ & $7.52 \pm 0.01$    & (2)   \\
                  & \multicolumn{2}{c}{$5.46 \pm 0.02$}  &       \\
$(B_T - V_T)$     & $0.71 \pm 0.01$ & $0.73 \pm 0.02$    & (2)   \\
$B$               & $6.20 \pm 0.01$ & $8.10 \pm 0.02$    & (E18) \\
                  & \multicolumn{2}{c}{$6.02 \pm 0.03$}  &       \\
$V$               & $5.56 \pm 0.01$ & $7.45 \pm 0.01$    & (E18) \\
                  & \multicolumn{2}{c}{$5.38 \pm 0.02$}  &       \\
$(B-V)$           & $0.64 \pm 0.01$ & $0.65 \pm 0.02$    & (E18) \\
                  & \multicolumn{2}{c}{$0.64 \pm 0.03$}  &       \\
$\Delta V$        & \multicolumn{2}{c}{$1.89 \pm 0.02$}  & (E18) \\
$M_V$             & \revone{$2.99 \pm 0.06$} & \revone{$4.88 \pm 0.06$}    & (E18) \\
\revone{$\Delta M_V$} & \revone{$2.51 \pm 0.06$} & \revone{$0.85 \pm 0.06$} & (E18) \\
$T_\eff$          & $5757 \pm 57$ K & $5705 \pm 73$ K    & (E18) \\
$L/\Lnom$         & \revone{$5.8 \pm 0.3$}   & \revone{$1.025 \pm 0.055$}  & (E18) \\
$R/\Rnom$         & \revone{$2.42 \pm 0.08$} & \revone{$1.04 \pm 0.04$}    & (E18) \\
$M/\Mnom$         & $1.70 \pm 0.64$ & $1.00 \pm 0.25$    & (3)   \\
                  & \revone{$1.58 \pm 0.26$} & \revone{$0.91 \pm 0.15$}    & (E18) \\ 
$P_\eq$           & \revone{$42 \pm 29$}    & --- & (E18)  \\
$\anglemean{P_\rot}$ & \revone{$40.2 \pm 2.3$ d} & \revone{$\sim 28$ d} & (4), (E18) \\
$\median{S}$ & \revone{$\sim 0.178$}     & \revtwo{$\sim 0.157$}       & (E18)  \\
$\log(\median{\RpHK})$ & \revone{$\sim -4.89$}     & \revone{$\sim -5.02$}       & (E18)  \\
\enddata
\tablerefs{
(1) \citet{ESA:1997} (2) \citet{Fabricius:2002}
(3) \citet{Pourbaix:2000} (4) \citet{Donahue:1993:thesis} and
  \citet{Donahue:1996} (E18) This work.
}
\tablecomments{Values between the A and B columns are the magnitude
  sums of A and B.  \revone{$\Delta M_V$ is the height above the main sequence
  as defined by \citet{Wright:2004}.  \revtwo{$P_\eq$ is the equatorial rotation based on $\vsini$ and assuming spin-orbit alignment.}  Values prefixed with $\sim$ are
  based on the assumed deconvolved activity scenario Case 1 from
  Section \ref{sec:deconv}.}}
\end{deluxetable}

Table \ref{tab:system} summarizes orbital properties and convolved
measurements for the HD 81809 system.  We take the latest orbital
solution updated using SOAR speckle photometry found in
\citep{Tokovinin:2015}, in particular the semimajor axis $a = 0.428
\pm 0.001$, orbital period $34.80 \pm 0.06$ yr, and inclination
$i_\orb = 85.4 \pm 0.1\degrees$.  This solution had a quality grade of
2 out of 5, which is characterized as ``Good: most of a revolution,
well observed, with sufficient curvature to give considerable
confidence in the derived elements.  No major changes in the elements
likely''  \citep{Hartkopf:2001}. \revone{We adopt the \emph{Gaia} DR2
  parallax $\pi = 29.63 \pm 0.76$ mas \citep{Gaia:2016,Gaia:2018:DR2}, which agrees to within the
  uncertainties with the earlier corrected
\emph{Hipparcos} parallax from \cite{Soderhjelm:1999} ($\pi = 29.0 \pm
1.1$ mas)}.  From this we obtain the mass sum from Kepler's third law, \revone{$M_A
+ M_B = (a/\pi)^3 / P_\orb^2 = 2.5 \pm 0.2 \, M_\Sun$}.  We adopt the radial velocity
measurements and resulting mass ratio $q = M_A/M_B = K_1/K_2 = 0.579
\pm 0.084$ of the ninth catalogue of
spectroscopic binary
orbits\footnote{\url{http://sb9.astro.ulb.ac.be/DisplayFull.cgi?1474+1}}
\citet[][hereafter $S_{B^9}$]{Pourbaix:2004}.  Combining these values, we obtain component
masses \revone{$M_A = 1.58 \pm 0.26 \, M_\Sun$ and $M_B = 0.91 \pm
  0.15 \, M_\Sun$}.  These more
precise values are consistent to within the uncertainties of the
earlier \citet{Duquennoy:1988} values used by \citet{Baliunas:1995}
for their interpretations of the binary.  However, the best estimate
mass of the A component has decreased and that of the B component has
increased, placing the latter closer to the solar mass and further
from the K-type star that was previously assumed.  Unfortunately, the
mass uncertainties remain large and are not particularly useful for
estimating the expected activity of the components.

%%%
\subsection{Tycho Photometry and Component Luminosity, Temperature, and Radius}

The Tycho Double Star Catalog
\citep[TDSC;][]{Fabricius:2002} contains resolved two-color photometry
for HD 81809 A \& B in the Tycho $B_T$ and $V_T$ bands, which are
close to the Johnson $B$ and $V$ bandpasses.  TDSC consists of a
re-reduction of the Tycho data, whose Tycho-2 catalog
\citep{Hog:2000} only contains resolved binary components for
separations of 0.8\arcsec{} or greater.  The double star
solution method for TDSC is close to that described in
\citet{Hog:2000}, but in this re-reduction solutions with separations
below the previous conservative limit of 0.8\arcsec{} were accepted.
The smallest accepted separation in TDSC is 0.29\arcsec{}, while the
reported separation for HD 81809 A \& B at that epoch was
0.52\arcsec{}.  The methods for the photometric solution are described
briefly in \citet{Fabricius:2000} and extensively in \citet{ESA:1997}.
Briefly, the Tycho photometry consists of averaging each slit crossing
of the target stars into a global average solution, where each
crossing consists of two series of 31 photon counts in the $B_T$ and
$V_T$ bands.  The TDSC photometry ($B_T$, $V_T$) and color index $B_T
- V_T$ for HD 81809 A \& B are given in Table \ref{tab:components}, as
well as their magnitude sums.  These magnitude sums may be compared to
the \emph{unresolved} Tycho-2 solution of \citet{Hog:2000} shown in Table
\ref{tab:system} as an internal consistency check.  The differences
(TDSC - Tycho-2) are $\Delta B_T = -0.025 \pm 0.033$, $\Delta V_T =
-0.023 \pm 0.022$, $\Delta (B_T - V_T) = -0.003 \pm 0.040$ 
indicating they are equivalent to within uncertainties.

We convert the TDSC photometry to the Johnson photometric system ($B$,
$V$) using the relationships presented in \citet{ESA:1997} (Sec 1.3,
Appendix 4, Equation 1.3.26).  The resulting $B$, $V$, and $(B-V)$ are
given in Table \ref{tab:components}, along with their magnitude sums.
As an external consistency check, we
compute the magnitude sums to the \citet{Johnson:1966} \emph{unresolved} values,
obtaining (TDSC - Johnson) differences of $\Delta B = 0.00 \pm 0.01$,
$\Delta V = 0.00 \pm 0.01$, and $\Delta (B-V) = 0.00 \pm 0.03 $.  This
\emph{exact} agreement is remarkable given the difficulty of resolving
these stars.  The resolved color indices are
$(B-V)_A = 0.64 \pm 0.01$ and $(B-V)_B = 0.65 \pm 0.02$, both of which
are equivalent to within uncertainties to the estimated solar
value, $(B - V)_\Sun = 0.653 \pm 0.003$ \citep{Ramirez:2012}.
However, using the system parallax we find that the absolute visual magnitude
of the A component, \revone{$M_{V, A} = 2.99 \pm 0.06$}, is well above
the main-sequence, while the B component, \revone{$M_{V, B} = 4.88 \pm 0.06$}, is
close to the solar value, $M_{V, \Sun} = 4.83$ \citep{Allen:1973}.

Using the \citet{Flower:1996} empirical bolometric corrections and
color-temperature relationship (tabulated in \citealt{Torres:2010}) and
the Stephan-Boltzman law we find the fundamental properties for the A
component to be $T_{\eff, A} = 5731 \pm 47$ K, \revone{$L_A/\Lnom = 5.8 \pm
0.3$, and $R_A/\Rnom = 2.42 \pm 0.08$}, where the units are
the nominal solar values recommended in \revone{IAU 2015 Resolution B3 \citep{Prsa:2016}}.  For the B component,
we obtain $T_{\eff, B} = 5705 \pm 73$ K, \revone{$L_B/\Lnom = 1.025 \pm 0.055$,
and $R_B/\Rnom = 1.04 \pm 0.04$}.  Here we again find the A
component to be evolved and the B component to be nearly identical to
the Sun within the uncertainties.  However, HD 81809 is relatively metal poor, with [Fe/H] = $-0.29$
(\citealt{Holmberg:2009}).\footnote{There are similar estimates from
e.g. \citet{Mishenina:2013} ($-0.28$), though some other estimates find even
lower metal content, e.g. \citet{Boeche:2016} ($-0.38$); \citet{Takeda:2005}
($-0.34$).  These estimates are based on blended spectra whose flux is
dominated by the A component.  It is unknown what effect the presence
of the B component may have on the metalicity estimates.}  Assuming
the metalicity estimate represents both components, we would expect
the metal-poor B component to be less luminous than the Sun if had
solar mass and age.

%%%
\subsection{Comparison with MESA Stellar Evolution Model}
\label{sec:mist}

\begin{figure}
  \label{fig:models}
  \centering
  \subfloat{
    \includegraphics[width=\linewidth]{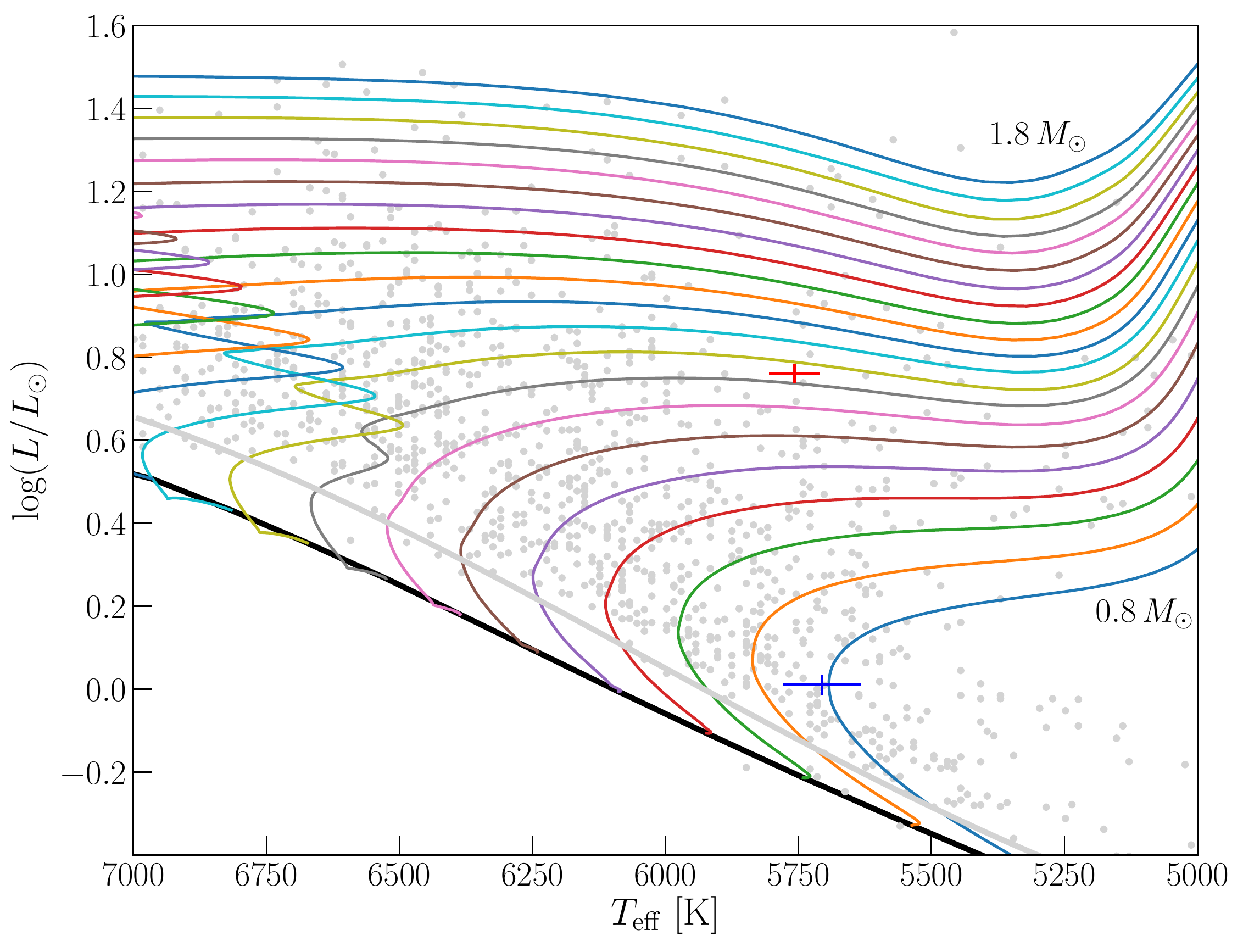}
    \label{fig:tracks}
  }
  \\
  \subfloat{
    \includegraphics[width=\linewidth]{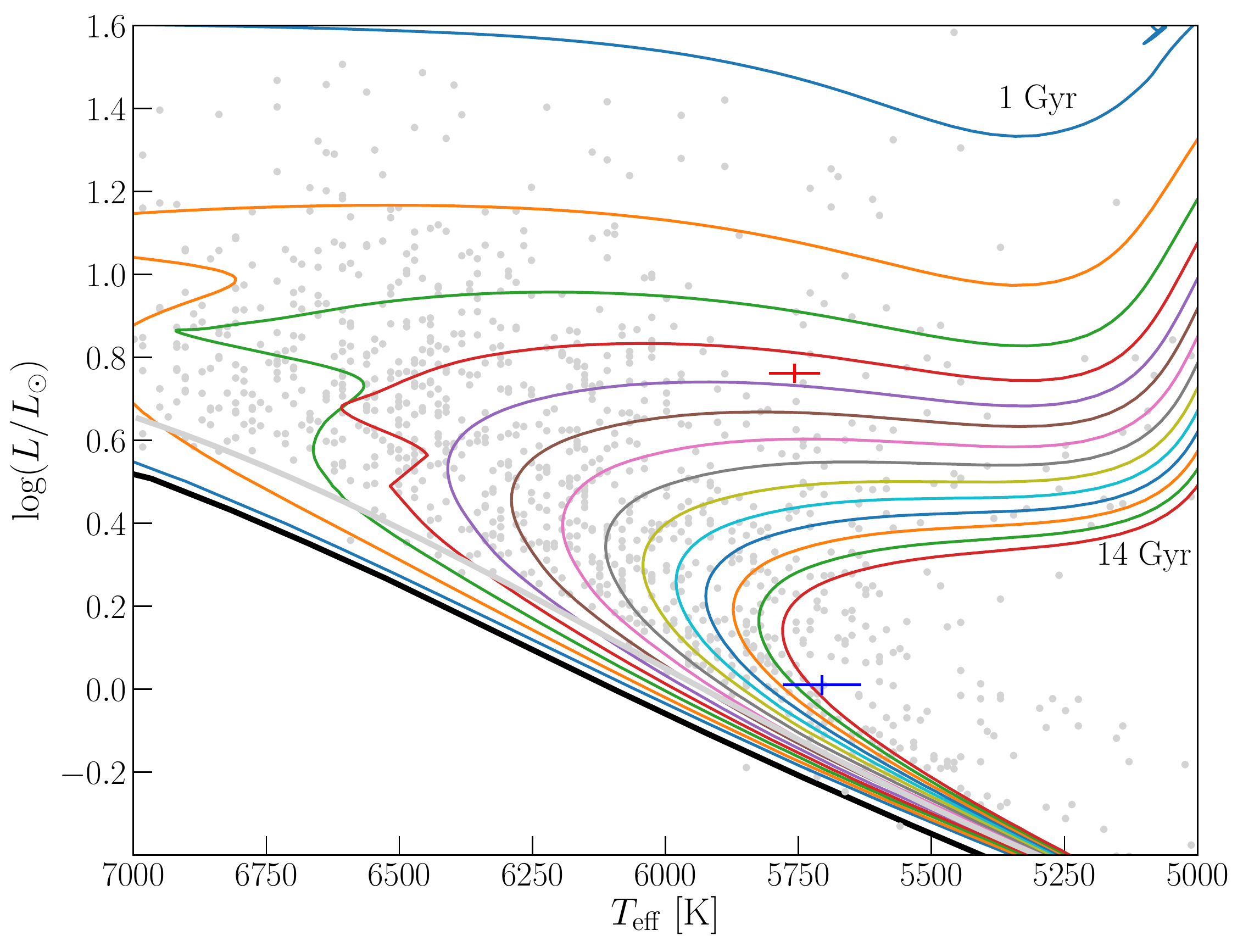}
    \label{fig:isos}
  }
  \caption{MIST evolutionary tracks and isochrones.  \emph{Top}:
    Colored lines are evolutionary tracks from $0.8 \, M_\Sun$ to $1.8
    \, M_\Sun$ in $0.05 \, M_\Sun$ increments with [Fe/H] = $-0.29$.  The
    black line is the zero-age main-sequence.  HD 81809 A (red) and B
    (blue) observations are overplotted.  Gray points are observations
    from the GCS with $-0.31 \leq [{\rm Fe/H}] \leq -0.27$, while the
    gray line is the estimated main-sequence fit to the lower edge of
    the observations.  \emph{Bottom:} Same as above, except that now
    the colored lines are model isochrones from 1 to 14 Gyr in 1 Gyr
    increments.}

\end{figure}

Stellar evolution models can help us check for consistency between the
the estimated mass (from orbital considerations) and the temperature
and luminosity (from photometry).  In Figure \ref{fig:tracks} we show
MESA Isochrones \& Stellar Tracks \citep[MIST;][]{Choi:2016}
evolutionary tracks from 0.8 $M_\Sun$ to 1.8 $M_\Sun$ interpolated to
[Fe/H] = $-0.29$ using their online
tool.\footnote{\url{http://waps.cfa.harvard.edu/MIST}} HD 81809 A
(red) and B (blue) are overplotted.  The model masses are about \revone{$1.18
\, M_\Sun$} for the A component and $0.81 \, M_\Sun$ for the B component.
These are lower than the masses estimated from orbital measurements by
1.5$\sigma$ and \revone{0.7$\sigma$}, respectively.  Furthermore, the B
component appears quite evolved, at the transition to the
main-sequence turnoff.  This indicates a great age for the B
component, which is confirmed the MIST isochrones plotted in Figure
\ref{fig:isos}, where the indicated age for the B component is over 14
Gyr.  This is older than the estimated age of the Universe.
Furthermore, A \& B do not lie on the same isochrone, as would be
expected for a co-evol binary.  These issues indicate that there is a
serious problem with either the characterization of HD 81809 A \& B or
the MIST models at this metalicity.  To check this, we have plotted a
selection of stars from the Geneva-Copenhagen Catalog (GCS;
\citealt{Holmberg:2009}) within 0.02 dex of the adopted metalicity.
These data reveal that dozens of stars are estimated to be older than
14 Gyr, and furthermore the lower edge of the GCS observations appears
to be higher than the model main-sequence.  To explore this, we
estimate the GCS main-sequence by fitting a third order polynomial to
the lowest luminosity stars in 50 K temperature bins (gray line).
Comparing this to the MIST model ZAMS, we find a non-uniform
difference with an average value in the domain [5300, 7000] K of 0.11
dex in $\log(L)$, or, switching the dependent variable, 207 K in
$T_\eff$.  The difference increases at higher temperatures, which may
indicate a scaling issue with the MIST models, or a bias to older
stars in the GCS data.

It is beyond the scope of this work to attempt to resolve the
difference in the main-sequence position between the MIST models and
the GCS observations.  We will only note that shifting the model grid
by the mean difference 0.11 dex places the A component at $\sim$6 Gyr
and the B component at \revone{$\sim$10 Gyr}, still about 2$\sigma$ away from
A's isochrone.  A larger shift of about \revone{0.16} dex is required to place
A \& B on the same $\sim$7 Gyr isochrone, however such an alteration
places a large number of low-temperature GCS observations below the
model main-sequence.  Such a deficiency may be in some sense
``normal'', as comparisons between MIST and well-characterized
eclipsing binaries shows the models are too luminous by $\sim$0.2 dex
for stars below $\sim$0.7 $M_\Sun$ \citep[see][Section 8.1]{Choi:2016}.  However,
shifting the grid by 0.17 dex further lowers the model masses to
$\sim$1.05 and $\sim$0.85 $M_\Sun$ for A \& B respectively.  This
places the mass sum about $1\sigma$ below the estimate based on
Kepler's third law.

A systematic underestimate of the HD 81809 metalicity or parallax
could also be a factor leading an overestimate of luminosities with
respect to the model grid.  Such errors would shift both the A \& B
components simultaneously, and so as above an \revone{$\sim$0.16} dex shift
would be required to place the components on the same model isochrone.
While an error in metalicity would only affect the model masses, a
larger parallax would also lower the orbit-based estimate of the
component masses so that the model and observations might ``meet in
the middle.''  We find that a parallax of $\sim$35\arcsec{} places both
components on the same MIST isochrone, however the resulting orbital B
mass becomes significantly smaller than the model mass.  Finally,
errors in the resolved two-color photometry would independently move
the A \& B components around the HR diagram.  From the above analysis
of the photometry we are encouraged that the TDSC measurements are
both internally and externally consistent with unresolved two-color
photometry on the Tycho and Johnson systems, respectively.
Furthermore, \citet{Fabricius:2002} Figure 8 compares the TDSC
photometry to the earlier reduction of \citet{Fabricius:2000}.  For
bright binaries like HD 81809, both the colors and the component $V_T$
magnitudes are quite consistent.  However, as the authors note, when
there are differences the TDSC reduction tends to be brighter, more so
for the B components.  Inspecting Figure \ref{fig:isos}, we see that
the B component would need $\Delta L \sim -0.2$ dex fainter to lie on
the same isochrone as the A component.  Using $\Delta M_\bol = \Delta
L/(-0.4)$, this corresponds to a brightness error of $+0.5$ mag.
Using the scatter in \citet{Fabricius:2002} Figure 8 as an rough
estimate of photometric precision, we can conclude that such a large
magnitude error would be very anomalous, but not impossible at the
brightness of HD 81809B.

To summarize, when comparing the HD 81809 component positions on the
HR-diagram to MIST model isochrones we find issues with the age of the
B-component, that it is (1) older than the Universe (2) not the same
age as the A-component, and furthermore (3) the model masses are lower
than masses estimated from the orbital parameters.  Adding GCS stars
of the same metalicity to the HR-diagram reveals that there may be a
model calibration issue whose resolution by a +0.11 dex shift in
luminosity may partially resolve issues (1) and (2), while worsening
problem (3).  A larger shift of +0.16 dex is required to fully resolve
age issues, while further worsening the mass issue.  An error in the
measured metalicity would have a similar conflict between model ages
and masses versus observations, while a larger parallax could help to
resolve most of the issues.  Errors in the resolved photometry are
still a possibility and may also play a factor.  Allowing for the
possibility of systematic luminosity calibration error in the model,
the model age for HD 81809 system should be in the range 4--7 Gyr, and
the corresponding model mass sum from 2.3--1.9 $M_\Sun$.

Due to the possibility for MIST model calibration errors, we will put the
model results aside and take the observations as they are for the
remainder of our analysis.

%%%
%%%
%%%
\section{Rotation-based Identification of the Active Component}
\label{sec:rot}

Because the cycle shown in Figure \ref{fig:timeseries} is so clean, we assume that
the variability is dominated by only one component in the
binary.  Similarly, we attribute the 40.2 day rotation measurement of
\citep{Donahue:1996} to the same active component.  We can
obtain an independent estimate of rotation period using the projected
rotational velocity measured from Doppler broadening, $\vsini$.

\begin{equation}\label{eq:Peqsini}
  \frac{P_\eq}{\sin i} = \frac{ 2 \pi R }{ v \sin i} ,
\end{equation}

\noindent where $P_\eq$ indicates the rotation period at the stellar
equator and $R$ is the radius of the star.  We use $v \sin i = 2.9 \pm 0.3$ km
s$^{-1}$ from \cite{AmmlerVonEiff:2012}.  Next, assuming the
rotational axis is aligned with the orbital axis, we use orbital
inclination $i = 85.4^\circ \pm 0.1^\circ$ from the binary solution of
\cite{Tokovinin:2015}.  This assumption is justified in the work of
\citet{Hale:1994}, who finds that in binary systems with separations $\log(a)
\lesssim 1.2$ ($\lesssim 16$ AU) the primary component equatorial plane and orbital
plane are co-planar to within $\pm 10\degrees$.  Using these values and the radius of the A
component from above in equation \eqref{eq:Peqsini} we find \revone{$P_{\rm eq,A} =
42 \pm 29$ d}.  The large uncertainty is dominated by the uncertainty in
$v \sin i$, however the best value is close to the measurement of
\citet{Donahue:1996}.  Since the Doppler broadening of the blended
spectra should be dominated by the much brighter A component ($\Delta
m_V = 1.89$ implies a flux ratio $F_A/F_B = 5.70$), and the rotation
period from the $S$-index record agrees with $P_{\rm eq}$, we conclude
that the $S$-index modulations of HD 81809 are dominated by the
evolved A component.  Using similar arguments for the B component, we
find $P_{\rm eq,B} = 18 \pm 13$ d, such that the angle between the orbital
axis and the rotational axis of B would have to be \revone{$59^\circ$} in order
for $P_{\rm eq,B}$ to agree with $P_\rot = 40.2$ d.  Such a large
spin-orbit misalignment may be difficult to explain on physical
grounds, however see \citet{Offner:2016}.

Furthermore, the long rotation period attributed to the A component is
consistent with our expectations from conservation of angular momentum
as an evolved star expands \citep{VanSaders:2013}.  Such a long rotation period for the
G-type main-sequence B component is inconsistent with the
observational findings of \citet{VanSaders:2016}, who found that
rotational braking ceases to operate past a critical Rossby number
($\equiv P_\rot/\tau_c$, where $\tau_c$ is the convective turnover
time; see \citealt{Noyes:1984}) of about 2, thereby \revone{temporarily} halting the growth of
the Rossby number until the star begins evolutionary expansion.  If
the B component were responsible for the 40.2 d rotational
modulations, its Rossby number would be 3.29.  The A-component
\citet{Noyes:1984} Rossby number is similarly large (3.56), however
the empirical relationship was developed using main-sequence stars,
and is likely invalid for evolved stars \citep[see][]{Gilliland:1985}.

%%%
%%%
%%%
\section{Deconvolving the $S$-index}
\label{sec:deconv}

%%%
\subsection{Mathematical Description}

Rewriting the $S$-index definition \eqref{eq:S} for an unresolved
binary with components A \& B, we have:

\begin{equation}
  S_{AB} = \alpha \frac{(N_{H, A} + N_{H, B}) + (N_{K, A} + N_{K, B})}{(N_{R, A} + N_{R, B}) + (N_{V, A} + N_{V, B})} .
\end{equation}

We now seek to rewrite this in terms of the $S$ indices of the
individual components, $S_A$ and $S_B$.  This can done with some
algebra using the definition of the $S$-index \eqref{eq:S} and
defining:

\begin{equation}
  \label{eq:D}
  D \equiv \frac{N_{R,A} + N_{V,A}}{N_{R,B} + N_{V, B}},
\end{equation}

\noindent resulting with:

\begin{equation}
  \label{eq:Sdeconv}
  S_{AB} = \frac{S_B + D S_A}{1 + D} .
\end{equation}

Notice that $D$ in equation \eqref{eq:D} is the ratio of the
denominators of the component $S$-indices, i.e. the photon counts in
the two pseudo-continuum bands from the components.
In developing an activity index based only on the flux in the H and K
bands, \citet{Middelkoop:1982} developed a color-correction factor related to
the $S$-index denominator:

\begin{equation}
  C_{\rm{cf}} \equiv (N_R + N_V) \gamma /f_\bol,
\end{equation}

\noindent where $f_\bol = \gamma 10^{-0.4(m_V + BC)}$ is the apparent bolometric
flux and $\gamma$ is a constant dependent on extinction.
Alternatively, we can relate the flux to the luminosity $f_\bol = L /
4 \pi d^2$, where $d$ is the distance.  Then we rewrite $D$ in equation \eqref{eq:D} as:

\begin{equation}
  \label{eq:Dmeas}
  D = \frac{L_A}{L_B} \cdot \frac{C_{\rm{cf}, A}}{C_{\rm{cf}, B}} .
\end{equation}

\citet{Middelkoop:1982} developed an empirical relation for $C_{\rm{cf}}$
as a function of $(B-V)$ (see their equation 8), and we have already
estimated the component luminosities above.  Therefore, $D$ is a known
quantity; for HD 81809 its value is 5.80.  Now if we can assume a value
for one of the component $S$-indices, then we can use the observed
convolved $S$-index $S_{AB}$ and equation \eqref{eq:Sdeconv} to solve
for the other component.

%%%
\subsection{Estimating Activity of the Flat Component}

As discussed above, we assume that only one of the components of HD
81809 is ``active'' and cycling, while the other has negligible
variability, or is a flat-activity star.  We can therefore use
observations of other flat-activity stars to estimate an $S$-index for
the flat component.  While \citet{Hall:2007b} finds flat-activity stars
at a large range of activity levels, most of those above
$\log(\RpHK) > -4.75$  were classified as variable in
\citep{Baliunas:1995} using longer time series.  \revone{We therefore
  use the long $\sim$50-year activity records
  from the \citet{Egeland:2017:thesis} sample of solar-analog stars to define a
  model for flat-activity on the main sequence.  The sample is defined
  by stars with overlapping observation records from both MWO and SSS
  with effective temperatures within $\sim$5\% of the solar value.
  Starting from the original 27-star sample, we measure the
  seasonal-median amplitude, defined as $A_s = \max(\median{S}_i) -
  \min(\median{S}_i$, where $\median{S}_i$ is the median
  $S$-index for season $i$.  We define ``low-variability'' as any star
  with $A_s$ lower than the solar value of 0.0203, finding 9 such
  stars. \revtwo{We define ``main-sequence'' following \citet{Wright:2004} using the 
absolute magnitude height 
  $\Delta M_V = M_{V, \rm{MS}}(B-V) - M_V$ above the
  \emph{Hipparcos} ensemble mean main sequence, $M_{V,\rm{MS}}(B-V)$, defined using the
eighth-order polynomial found in the erratum of \citet{Wright:2004}.
Stars with $\Delta M_V < 1$ are considered ``main-sequence'', and
those with larger $\Delta M_V$ are ``evoloved.''  This $\Delta M_V$ cut leaves us
  with 7 stars} with effective temperatures within 2\% of the solar
  (and HD 81809 A or B) value.  The main-sequence flat-star sample properties are
  shown in Table \ref{tab:flatstars}.  The $A_s$ for this sub-sample
  ranges from \revtwo{0.0085--0.0178, or 41--87\% of the solar variability.
  The median activity ranges from $\median{S}=$ 0.147--0.179, with
  an ensemble median of \revtwo{0.157}} (or $\log(\RpHK) = -5.02$ using $(B-V) =
  0.65$).  We use this sub-sample ensemble median activity to
  represent the ``typical'' activity of a low-variability
  main-sequence solar-analog star.}

\begin{deluxetable}{lcccccc}
\tablecaption{Low-Variability Main-Sequence Stars\label{tab:flatstars}}
\tablehead{
  \colhead{HD} &  \colhead{$T/\Tnom$} &  \colhead{$R/\Rnom$} &
  \colhead{$L/\Lnom$} & \colhead{\revtwo{$\Delta M_V$}} & \colhead{$\median{S}$} & \colhead{$A_s$}
}
\startdata
43587  & 1.02 & 1.24 & 1.62 & 0.77 & 0.1552 & 0.0111 \\
71148  & 1.00 & 1.11 & 1.24 & 0.72 & 0.1567 & 0.0145 \\
126053 & 0.98 & 0.92 & 0.79 & 0.22 & 0.1663 & 0.0134 \\
141004 & 1.02 & 1.38 & 2.07 & 0.92 & 0.1573 & 0.0178 \\
142373 & 1.01 & 1.73 & 3.06 & 0.89 & 0.1470 & 0.0085 \\
143761 & 1.00 & 1.31 & 1.74 & 0.71 & 0.1505 & 0.0143 \\
176051 & 1.01 & 1.20 & 1.51 & 0.46 & 0.1793 & 0.0157 \\
%217014 & 0.99 & 1.19 & 1.37 & 0.1493 & 0.0118 \\
\enddata
\end{deluxetable}

\citet{Wright:2004} examined the previous suggestions that stars with
very low activity ($\log(\RpHK) <= -5.1$) may be in suppressed
activity states analogous to the solar Maunder Minimum.  They found
that most stars in that activity range were evolved \revtwo{according
  to the definition given above.}  More strictly,
they found that \emph{all} such evolved stars had $\log(\RpHK) <
-4.9$, while $\log(\RpHK) ~ -5.1$ is approximately the median activity
of evolved stars (see their Figure 3).  Therefore, if the evolved A
component were the flat-activity component, $\log(\RpHK) \sim -5.1$ is
a reasonable estimate for its activity.

\begin{figure}
  \centering
  \subfloat{
    \includegraphics[width=\linewidth]{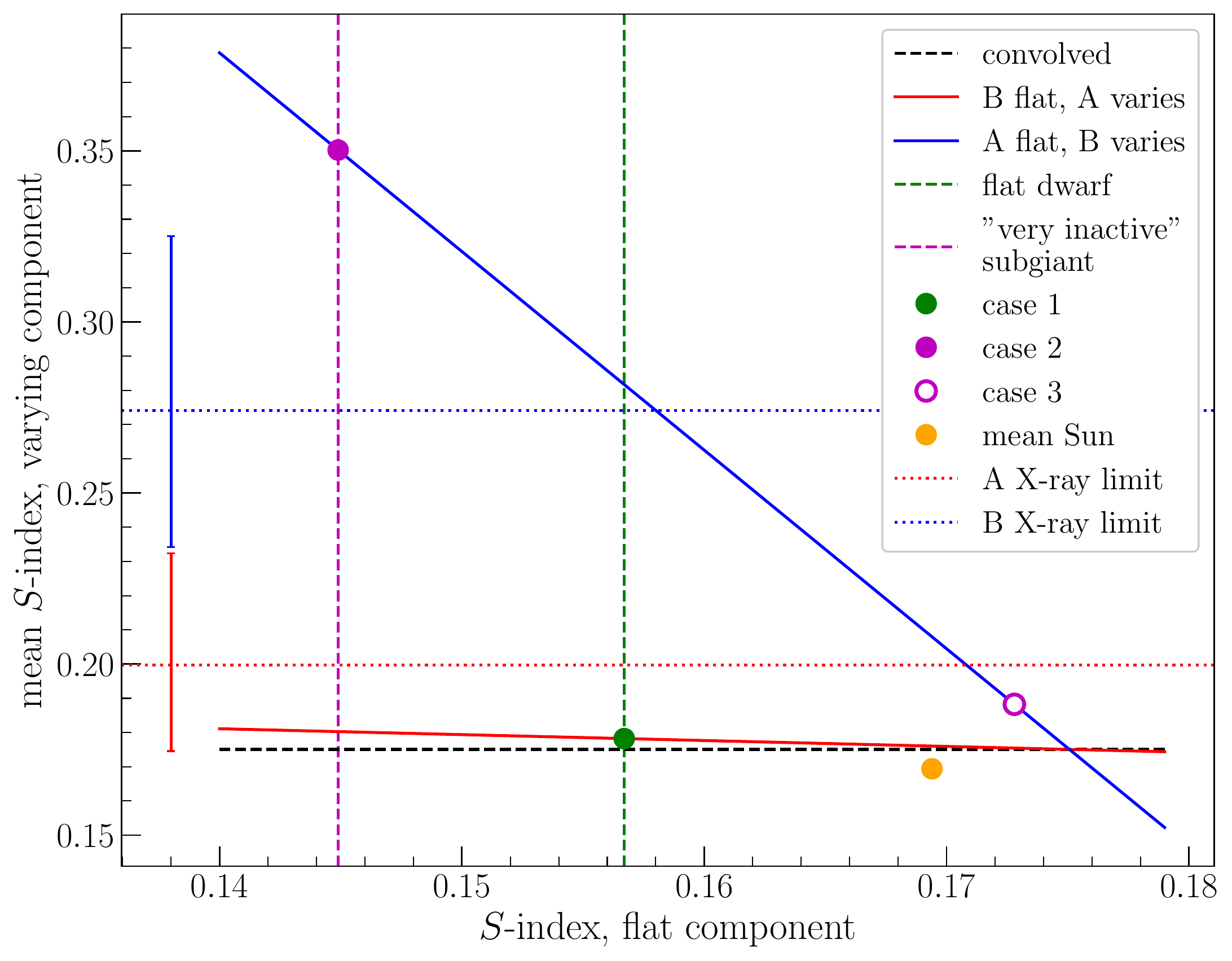}
    \label{fig:deconv_S}
  }
  \\
  \subfloat{
    \includegraphics[width=\linewidth]{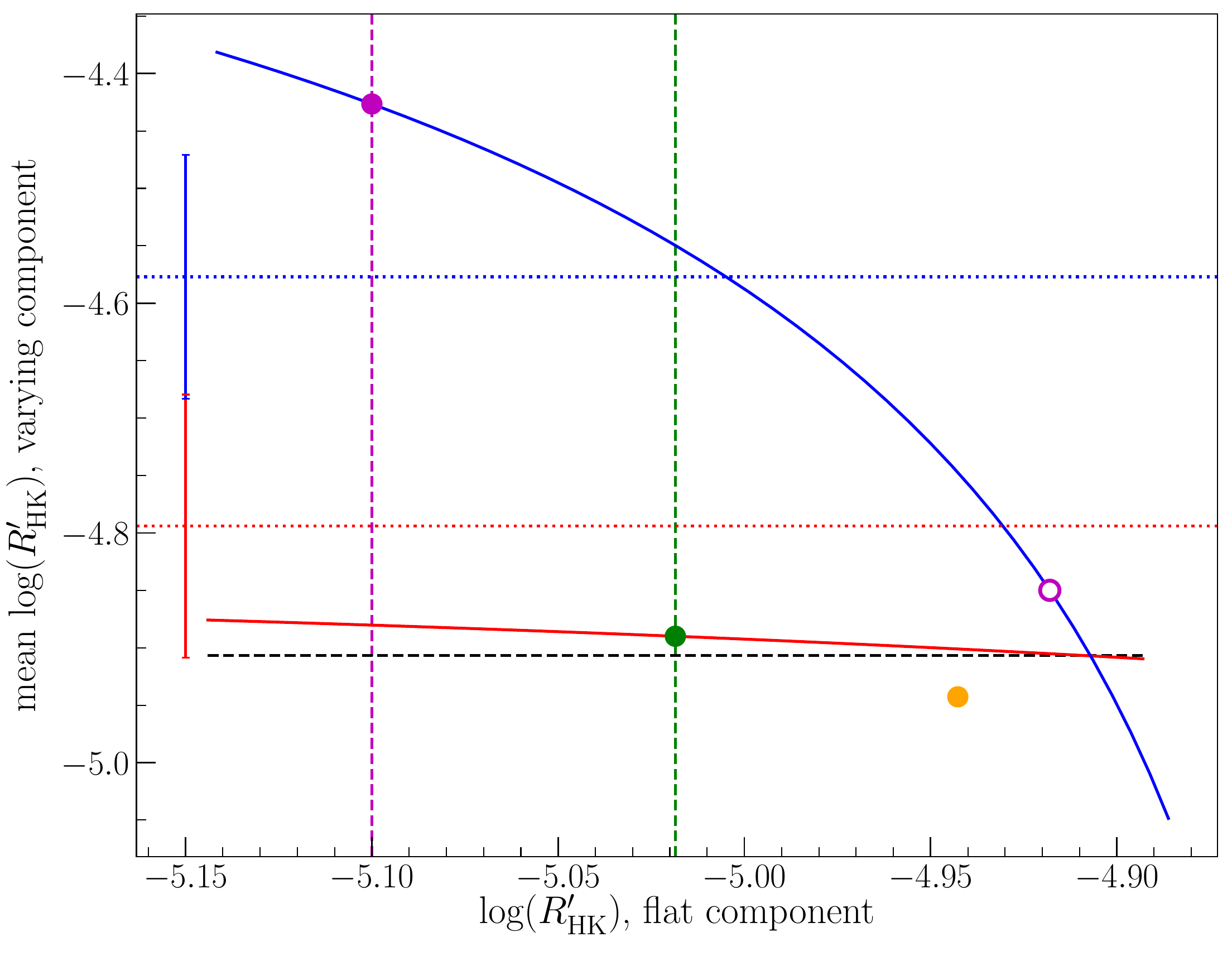}
    \label{fig:deconv_logRpHK}
  }
  \caption{Deconvolving the $S$-index.  \emph{Top}: $S$-index
    deconvolution where the x-axis is the assumed activity of the flat
    component of HD 81809 and the y-axis is the resulting mean
    activity of the varying component.  The convolved $S$-index is
    shown as a black dashed line for reference.  The red line comes
    from equation \eqref{eq:Sdeconv} and the assumption that the
    secondary is flat, while the blue line assumes that the primary is
    flat.  Vertical dashed lines give best estimates for the likely
    activity of a flat-activity dwarf (green) and subgiant (magenta)
    respectively.  Colored circles give the HD 81809 decomposed
    activity cases discussed in the text, with the mean solar activity
    (orange) plotted for reference.  \revone{Horizontal dotted lines
    and associated error bars give an upper limit to activity
    assuming all of the measured X-ray flux
    comes from only one component.} \emph{Bottom}: Same as above, but
    using the $\log(\RpHK)$ activity index.}
  \label{fig:deconv}
\end{figure}

In Figure \ref{fig:deconv} we use equations \eqref{eq:Sdeconv} and
\eqref{eq:Dmeas} along with the component data in Table
\ref{tab:components} to model the component $S$-indices for various
assumed activity levels of the flat component.  Three scenarios are
shown.  In Case 1, the luminous subgiant A is the variable component
and the flat activity for B is assumed to be \revone{$S \sim 0.157$,
  estimated using the low-variability solar-analog main-sequence
  sub-sample described above.}  In Case 2, the main-sequence B
component is variable and the subgiant A has an assumed $\log(\RpHK)
\sim -5.1$, estimated using the \citet{Wright:2004} data.  Finally,
Case 3 again assumes B is variable and A is flat, but fixes the
activity of B to a value of $\log(\RpHK) = -4.85$ that would not be
anomalous on the activity-rotation diagram (see below).  In the lower
panel $S$ is transformed into $\log(\RpHK)$ using the formulation of
\citet{Noyes:1984} and the $B-V$ colors of Table \ref{tab:components}.
The mean solar activity ($S_\Sun = 0.1694; \log(R'_{\rm HK, \Sun}) =
-4.9427$) from \citet{Egeland:2017} is shown for reference.

Figure \ref{fig:deconv} shows that when the less-luminous B component
is assumed to be flat, as we believe based on the rotational analysis
of Section \ref{sec:rot}, then the activity of the varying component
$S_A$ is not much different than that of the original convolved
observation regardless of the activity level of the B component.  \revone{In
Case 1, using our assumed $S_B \sim 0.157$, we would find $S_A =
0.178$, slightly more active than the Sun.  Converting to $\RpHK$, we
have $\log(R'_{\rm HK, B}) \sim -5.02$ implying that $\log(R'_{\rm HK,
  A}) = -4.89$.}  Assuming a lower $S_B$ results in nearly negligible
increases in $S_A$, due to the small slope $\Delta S_A/\Delta S_B =
0.172$.

However, if the more-luminous A component is assumed to be flat, the
activity of the B component $S_B$ is quite large at the ``typical''
activity of a subgiant.  Case 2 assumes $\log(R'_{\rm HK, A})
\sim -5.1$, resulting in $\log(R'_{\rm HK, B}) =
-4.43$ (in $S$, $S_A \sim 0.145$ and $S_B = 0.347$).  Significantly
higher activity in A would be required to push B into a low-activity
regime, which would be required to keep its activity from being
anomalous with respect to its rotation (see Figure \ref{fig:ensemble}
below).  Such a situation is given in Case 3, where we fix the B
activity to $\log(\RpHK) \sim -4.85$ and solve for the required
activity of A, obtaining $\log(\RpHK) = -4.92$.

%%%
\subsection{Limits Imposed by X-ray Observations}

\revone{
We used the X-ray/calcium HK relationship described in the appendix of
\citet{Mamajek:2008} to check whether our deconvolved activity levels
are compatible with the convolved X-ray observations for HD 81809.  We
compute upper limits to the expected $\log(\RpHK)$ emission by
assuming that \emph{all} of the X-ray emission comes from either the A
or the B component.  \citeauthor{Mamajek:2008}'s relationship is based
on $R_X = L_X / L_{\rm bol}$, which we computed using the data from
Table \ref{tab:components}.  Then, assuming all X-ray emission is from
A, we would expect $\log(R'_{\rm HK, A}) \leq -4.79 \pm 0.11$.
Similarly for the B component we would expect $\log(R'_{\rm HK, B}) \leq -4.58 \pm
0.11$.  The uncertainties given are based purely on the parameter
uncertainties of the empirical relationship of \citet{Mamajek:2008}.
These upper limits are plotted on Figure \ref{fig:deconv} along with
the deconvolved activity curves and the three special cases described
above.  The entire activity curve for scenarios in which B is flat and
A is the varying component is allowed by the X-ray upper limit,
including our best estimate Case 1.  For scenarios in which A is flat
and B is varying, the X-ray data renders the high-activity portion of
the curve unlikely, including Case 2 that assumed the subgiant A
component has the typical value from \citet{Wright:2004}, $\log(\RpHK)
= -5.1$.  The X-ray data do not fully exclude Case 2, which lies just
over 1$\sigma$ above the upper-limit estimate.  Case 3, which is
chosen such that the B component does not have an anomalously high
activity for a $\sim$40 day rotation period (see below), is found to
be fully compatible with the X-ray upper limit.
}

%%%
\subsection{Comparison to Activity--Rotation Data}

\begin{figure}
  \centering
  \subfloat{
    \includegraphics[width=\linewidth]{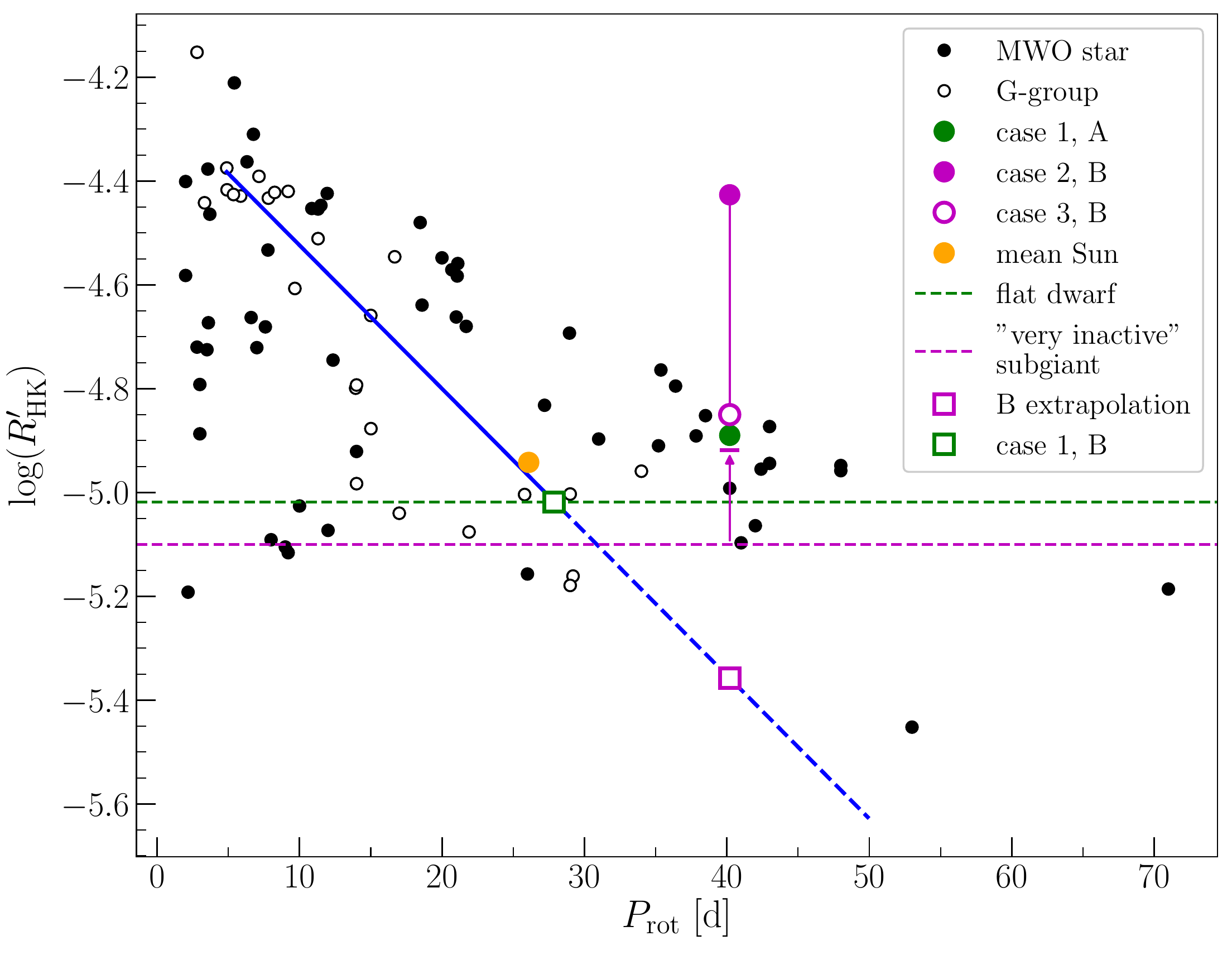}
    \label{fig:ensemble_Prot}
  }
  \\
  \subfloat{
    \includegraphics[width=\linewidth]{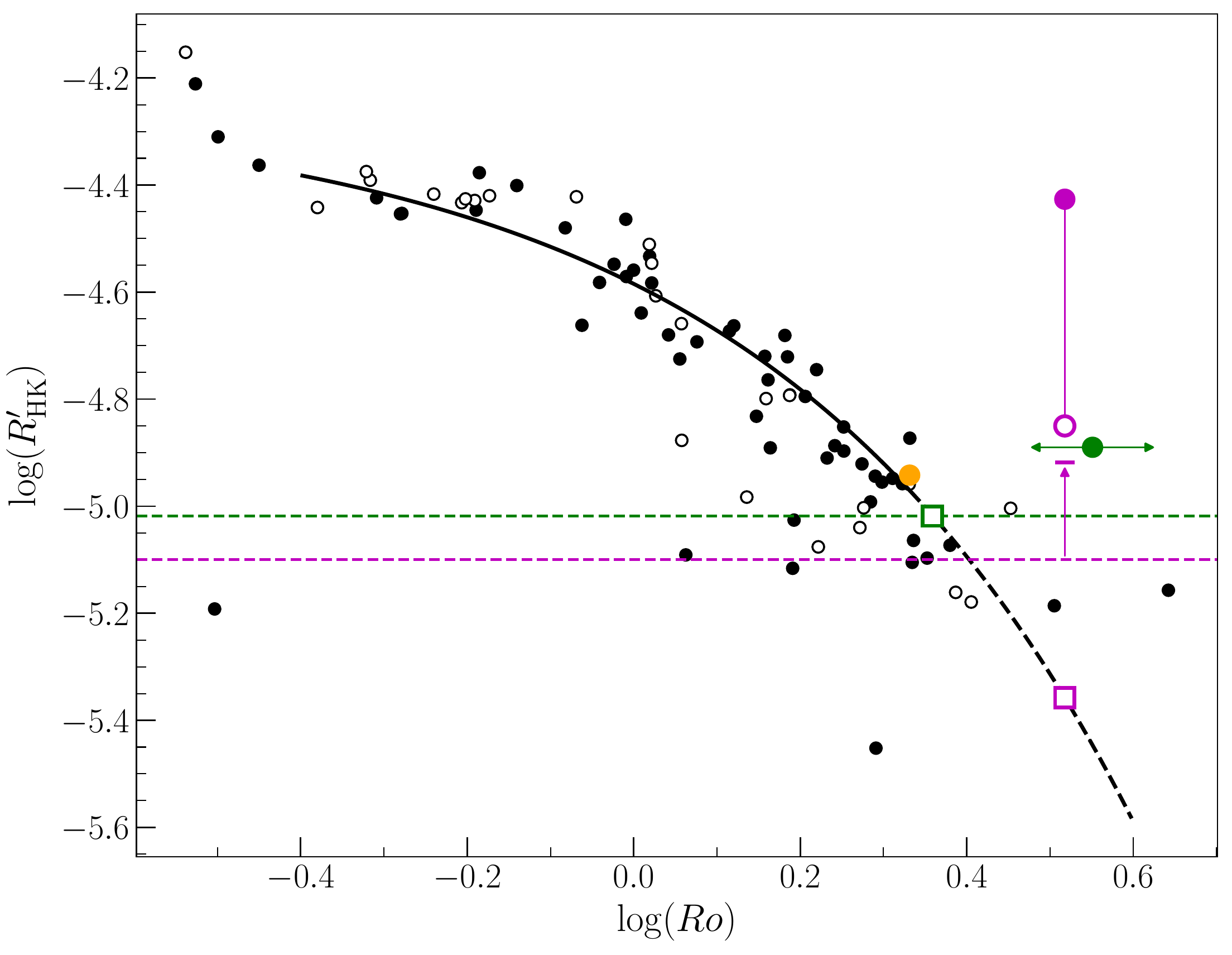}
    \label{fig:ensemble_Ro}
  }
  \caption{Deconvolved HD 81809 activity-rotation (Top) and
    activity-Rossby (Bottom) cases from Figure \ref{fig:deconv}
  compared to the MWO ensemble with measured rotation periods (black
  filled circles; G-type stars are open circles).  \revone{In the top
    panel, the blue line is the activity-rotation relationship from
    \citet{Mamajek:2008} for the B component, while in the bottom
    panel the general activity-Rossby
    relationship is shown. In both curves the dashed continuation shows an
    extrapolation for $\log(\RpHK) < -5.0$ that was not considered in their
  analysis.}  The colored circles
  and lines have the same meaning as in Figure \ref{fig:deconv},
  \revone{with the addition of the magenta square representing the
    expected activity level of the B component if it were the
    ``varying'' component and also followed the
    extrapolated \citet{Mamajek:2008} activity-Rossby relationship,
    and the green square showing the expected rotation of the
    B-component given the prototype ``flat dwarf'' activity level of
    Case 1.}
  Arrows indicate the directions of possible activity or Rossby number
  corrections.}
  \label{fig:ensemble}
\end{figure}

\revone{We now compare the three scenarios discussed
  above} with an ensemble of stars with known activity and
rotation.  We use the Mount Wilson sample of 82 stars with rotations
from the literature \citep{Egeland:2018b} for this purpose.  Figure
\ref{fig:ensemble} shows the HD 81809 activity scenarios with the MWO
ensemble.  We see from the figure that for our best estimate of a case
where A is varying and B is flat (Case 1) the HD 81809
point falls within the ensemble bounds for activity and rotation.
However, for our best estimate with B varying and A flat (Case 2), B
would be a severe outlier with respect to the ensemble.  Case 3 is an
upper limit for activity for which a varying B component would
\emph{not} be anomalous \revone{with respect to the ensemble.}  In this case, a more active A component is
required, as indicated by the magenta arrow.  Continuing this trend,
it would in fact be possible that the mean activity of a variable B
component is \emph{smaller} than a flat A component, which would not
necessarily place A as an outlier with respect to the ensemble, since
\revone{in this case its rotation is unknown and it may be fast
enough to avoid being anomalous.}

\revone{
Figure \ref{fig:ensemble} also shows the empirical activity-Rossby number relationship
of \citet{Mamajek:2008} (equation 6).  This relationship is based
on data with $-5.0 \leq \log(\RpHK) \leq -4.3$, but for illustrative purposes we
include the lower-activity extrapolations as dashed lines.  In the top panel of Figure
\ref{fig:ensemble} the rotation curve specifically refers to the B
component ($P_\rot = \tau_c \cdot Ro$, where $B-V = 0.65$ results in
$\tau_c = 12.2$ days using the relationship of \citealt{Noyes:1984}).  The
magenta square shows the extrapolated activity level $\log(\RpHK) =
-5.36$ if the B component were the varying component with $P_\rot =
40.2$ as derived from the Ca HK time series.  Such a low activity appears
to be unphysical in light of the large ensemble of Figure
\ref{fig:dMag} below.  The green square shows the expected rotation period
$P_\rot \sim 28$ days for the B component under the ``flat dwarf'' activity scenario
of Case 1.  This estimate appears to be reasonable by analogy; it is
not far below the Sun in parameter space and two other G-type stars
have similar activity and rotation measurements.  Note that we cannot
similarly estimate the expected rotation of the A-component under the
Case 2 and Case 3 scenarios because the \citet{Mamajek:2008}
activity-Rossby relationship is valid only for main-sequence stars.
}

In all cases the computed Rossby number \citep[using][]{Noyes:1984} for the varying
component is large with respect to the ensemble.  However, if the
varying component is A (Case 1), the convective turnover time may in fact be
in error due to the evolution \citep[see][Figure 10]{Gilliland:1985}.  The
estimated effective temperature of the A component lies almost exactly
on a transition point in Gilliland's models, where lower temperatures
lead to a steep increase in convective turnover time with respect to the
main-sequence value at that temperature, while higher temperatures
have a steep relative decrease.  Therefore, the correction for A could go
either way, as indicated by the green arrows.  The B
component, on the other hand, is close to the main sequence and so its
Rossby number estimation should be as good as any other member of the
ensemble.  Case 2 would place the B component activity in a highly
anomalous position, while Case 3 would be less so given the large
scatter in Rossby number at low activities.

The activity decomposition above does not conclusively decide whether
A or B is the varying component.  However, the case where A is varying
and B is flat results in deconvolved activity levels that are not
anomalous with respect to rotation for any reasonable deconvolution
scenario.  On the contrary, using the best estimate of a mean activity
level for a flat subgiant (Case 2) results in an anomalously high
activity level for a B component with a $\sim$40 d rotation period.
If one assumes a higher activity ($\log(\RpHK) > -4.92$) for a flat
subgiant A component this problem can be resolved (e.g. Case 3).  It
remains uncertain whether such an object exists in nature.  In the
\citet{Baliunas:1995} sample, highest activity luminosity class IV or
IV-V star classified as ``flat'' is 31 Aquilae (HD 182572) with
$\log(\RpHK) = -5.097$ \citep{Egeland:2018b}.

%%%
\subsection{Comparison to Activity--Evolution Data}

\begin{figure}
  \centering
  \includegraphics[width=\linewidth]{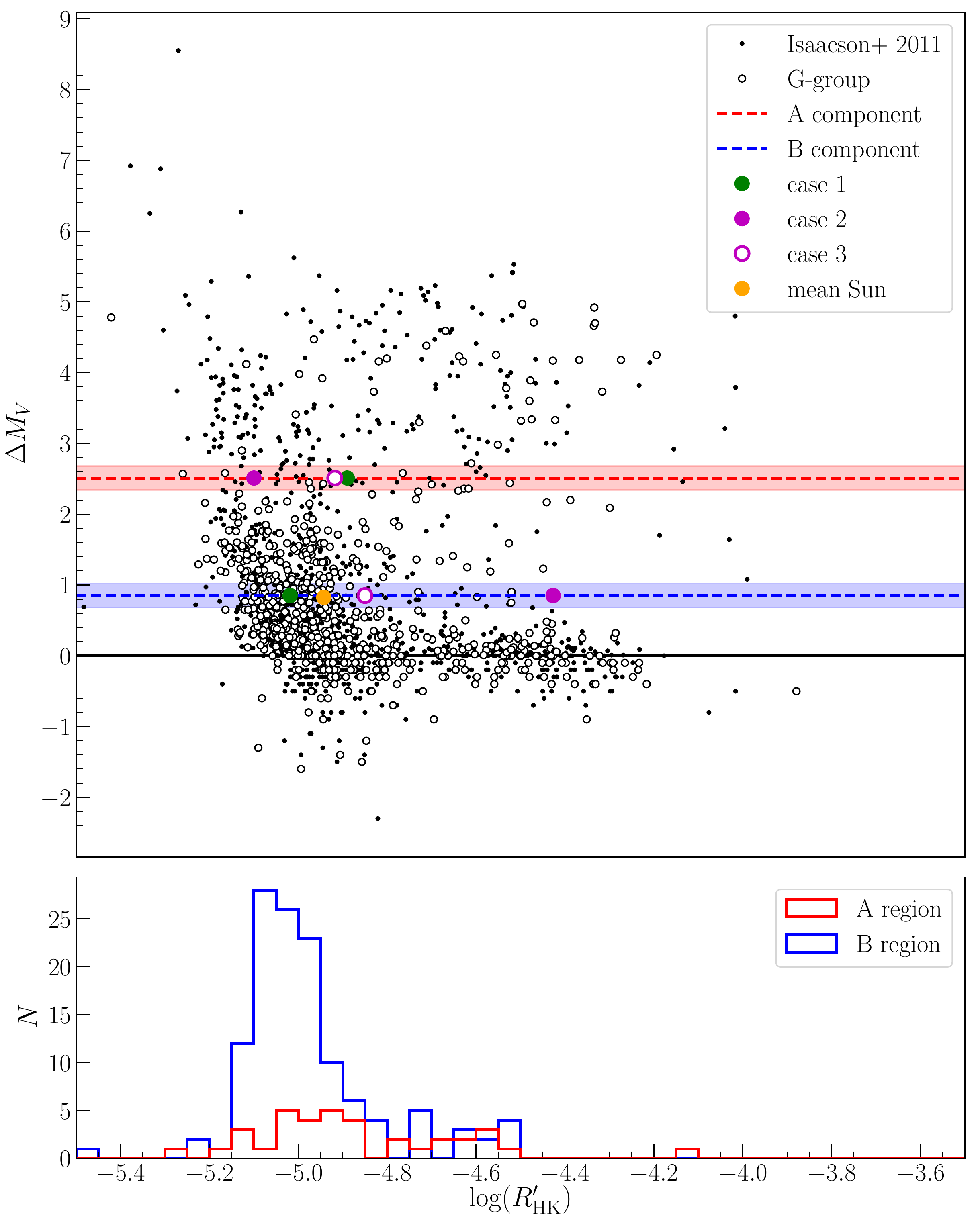}
  \caption{\revone{
    Height above the main sequence $\Delta M_V$ versus activity for
    the catalog of \citet{Isaacson:2010} with $0.44 \leq (B-V) \leq
    0.90$. G-type stars ($0.60 \leq (B-V) \leq 0.70$) are shown as
    yellow points.  The $\Delta M_V$ of the A and B components of HD 81809 are
    indicated with red and blue lines, respectively, with the shaded regions
    indicating the $3\sigma$ uncertainty.  The three cases of
    deconvolved activity given in Figure \ref{fig:deconv} are
    shown as colored circles.  Histograms
    of the data contained in the $\pm 3 \sigma$ shaded regions are
    shown in the bottom panel.
  }}
  \label{fig:dMag}
\end{figure}

\revone{
We examine whether the three activity scenarios derived above are
consistent with the observed activity levels from an ensemble of stars at
the same point in their evolution as HD 81809 A and B.
\revtwo{Following \citet{Wright:2004}, we compute the height above the
  \emph{Hipparcos} main sequence $\Delta M_V$} for the activity sample of
\citet{Isaacson:2010} with $0.44 \leq (B-V) \leq 0.90$ (1764 stars).
The results are shown in Figure \ref{fig:dMag}, which is an updated
version of Figure 7 from \citet{Wright:2004}.  Examining Figure
\ref{fig:dMag}, we see that for the subgiant A component any of the
three cases outlined above is compatible with the ensemble behavior,
although there are notably few stars around $\Delta M_{V, A} = 2.51$.  However,
for the dwarf B component ($\Delta M_{V, B} = 0.85$) Case 1 is
the most likely in a statistical sense, while Case 3 is a less likely
possibility.  Case 2 places the B component in an outlying region
which appears to be highly unlikely in the \citet{Isaacson:2010}
ensemble.  \revtwo{Both Case 1 and Case 3 place the B component very near to
the Sun ($\Delta M_V = 0.83$) in the activity-evolution plane.}
}

%%%
\subsection{Inclination and Differential Rotation}
\label{sec:inc_diffrot}

\revtwo{
\citet{Donahue:1996} measured seasonal rotation periods for HD 81809
ranging from 37.0 to 43.0 days.  Assigning these periods to the A component
and using the $\vsini$ and radius of Table \ref{tab:components}, we
can estimate some limits on the inclination of the rotation axis,
$i_\rot$, and the nature of its differential rotation.  Recall that assuming
spin-orbit alignment we obtained $P_\eq = 42 \pm 29$ days.  Ignoring
for the moment the enormous uncertainty, we can see that if HD 81809A
is spin-orbit aligned, then it must have anti-solar or perhaps banded,
Jupiter-like differential rotation, as the mean rotation period
$P_\rot = 40.2$ d is faster than the required equatorial rotation.
Conversely, if we assume strict solar-like differential rotation
(monotonically increasing rotation period with increasing latitude) and
assign the shortest seasonal rotation period to the equator, we find
$\sin(i_\rot) = 0.88 \pm 0.68$, or  $i_\rot = 61\degrees \pm 111\degrees$, for a spin-orbit misalignment of $\sim
24\degrees$, significantly larger than the solar misalignment of
$7.25\degrees$ \citet{Allen:1973}.  Given the large (indeed,
geometrically impossible) formal uncertainties in the above scenarios
no strong conclusions can be drawn.  However, there is some reason to
suspect that the formal uncertainties in inclination, which are
dominated by the uncertainty in the $\vsini$, are too large.
\citet{Hale:1994} was able to use similar inclination calculations to
find a distribution of spin-orbit misalignments smaller than $\sim 10
\degrees$ for binaries with orbital separations $\lesssim 16$ AU.
This result should not have been possible if the inclination
uncertainties cover the entire domain $[0\degrees, 90\degrees]$ as in the above case.  It also bears
noting that many of \citeauthor{Hale:1994}'s $\vsini$ measurements had
as large or worse relative uncertainty as our adopted $\vsini$ for HD
81809.  }

%%%
%%%
%%%
\section{Conclusions}
\label{sec:conclusions}

\revone{
The latest visual \citep{Tokovinin:2015} and radial velocity
\citep{Pourbaix:2004} orbital solutions, along with the \emph{Gaia}
DR2 parallax for the HD 81809 system indicates component masses of
$M_A = 1.58 \pm 0.26 \, M_\Sun$ and $M_B = 0.91 \pm 0.15 \, M_\Sun$.
Assuming the Tycho resolved two-color photometry and the \emph{Gaia}
parallax are accurate, we conclude that the A component is a subgiant
($T_\eff = 5757 \pm 57$ K, $L/\Lnom = 5.8 \pm 0.3$), the B component
is a main-sequence dwarf ($T_\eff = 5705 \pm 73$ K, $L/\Lnom = 1.025
\pm 0.055$), and that it is the subgiant A component that is
responsible for the well-defined 8.17 yr magnetic activity cycle and
40.2 d rotation period found in archival \emph{convolved} Ca HK
observations.  The latter conclusion is based on the following lines
of evidence:

\begin{enumerate}
  \item An independent estimate of rotation period using the convolved
    $\vsini$ measurement dominated by the larger flux of the A
    component, along with the A component's radius, results in a value
    of $\sim 40$ d that is consistent with the rotation period
    measured from the convolved Ca HK observations.  Doing a similar
    exercise for the B component results in disagreement.
  \item By assuming a typical value for the Ca HK activity of a flat,
    main-sequence B component, we derive an activity for the A
    component that is not anomalous with respect to ensemble
    activity-rotation data or the expected activity derived from the
    X-ray flux, while doing the converse and assuming a typical
    activity for a flat A component results in anomalous activity for
    the B component (see Figures \ref{fig:deconv} and
    \ref{fig:ensemble}).
  \item The activity scenarios for a cycling B component and flat A
    component result in atypical activity for the B component with
    respect to similarly evolved stars (see Figure \ref{fig:dMag}).
\end{enumerate}

Based on analogy with ensemble values, we find $\log(R'_{\rm HK, A}) \sim
-4.89$ at $P_{\rot, A} = 40.2 \pm 2.3$ d and $\log(R'_{\rm HK, B}) \sim
-5.02$ at $P_{\rot, B} \sim 28$ d to be a reasonable model for
deconvolved activity and rotation.  Resolved activity measurements are
necessary to characterize the system with certainty.  The Sun-like
magnetic activity cycle for a slowly rotating but significantly more luminous
star with a Sun-like effective temperature represents a new challenge
for dynamo theory.
}
%%%
%%%
%%%
\section{Discussion}
\label{sec:discussion}

\revone{Our conclusions depend upon the accuracy of the spatially
resolved Tycho two-color photometry as well as the the (however
reasonable) assumption that the HD 81809A rotation axis is parallel
with the orbital axis.  The Tycho resolved photometry is near the
edge of the instrument's capability \citep{Fabricius:2002},} but the
observations are at least consistent with earlier convolved
photometry, and the deconvolved activity level for a varying A
component is not anomalous with respect to ensemble activity-rotation
trends while the contrary case can be.  \revone{Future data releases
from \emph{Gaia} should be able to confirm the color temperature of
the HD 81809 components using the $G_{\rm{RP}}$ and $G_{\rm{BP}}$ bands of
that instrument.}  However, conclusive evidence of an active A
component can only come from resolved spectral observations of an
established proxy for magnetic activity.  Such observations should
be possible with modern telescopes with adaptive optics and
favorable conditions.

Our estimated Ca \II{} H \& K activity level for the A component is
\revone{near the middle of the activity distribution for evolved stars with
$\Delta M_V > 2$ in Figure \ref{fig:dMag}}, while the highest activity
evolved flat star in \citet{Baliunas:1995} has \revone{a lower
  activity,} $\log(\RpHK) = -5.097$.
We therefore speculate that perhaps all of the relatively
high-activity evolved stars are cycling or otherwise variable, while
those with lower activities may indeed be flat.  HD 81809A would not
be the first cycling subgiant reported.  Notably, HD 78366 (G0IV-V)
and HD 219834A (G8.5IV) had ``good'' cycles in the
\citep{Baliunas:1995} study, and five other evolved stars had
lower-quality cycles reported \citep{Egeland:2018b}.  The quality of
the HD 81809 cycle is exceptional, however, and in that regard
remarkable for a subgiant.  The MWO HK Project began to
\emph{purposefully} observe evolved stars in the early 1990s and
preliminary results were given in \citet{Baliunas:1998}.  It was shown
that the same variety of patterns in long-term variability that exist
for main-sequence stars also exist in evolved stars.  Unfortunately,
the detailed study was never published and observations ceased in
2003.  More long-term variability studies are necessary to determine
patterns of magnetic activity with evolution.  In particular, the
widely-used Rossby number formulation of \citet{Noyes:1984} likely
does not work for evolved stars \citep{Gilliland:1985}.  A corrected
approach may result in reduced scatter in activity-Rossby
distributions (e.g. Figure \ref{fig:ensemble}) at the low-activity end
where most of the stars are evolved \citep{Wright:2004}.  Based on the
work of \citet{Gilliland:1985}, it is indeed possible that the Rossby
number of HD 81809A is rather close to the solar value, which could
make it an unlikely kind of ``solar twin'' in all the ways
which matter to the dynamo \citep{Egeland:2018b}.

\revtwo{
In Section \ref{sec:inc_diffrot} we found that if HD 81809A is
spin-orbit aligned, then it must have anti-solar or Jupiter-like
banded differential rotation, or conversely if it has strictly
solar-like differential rotation then it must have a large spin-orbit
misalignment of 24\degrees.  We caution, however, that these
conclusions can only be drawn by ignoring the large formal
uncertainties on the derived inclinations, which are primarily due to
the large uncertainty in $\vsini$.  A scenario of solar-like
differential rotation and large spin-orbit misalignment is contrary to
the results of \citep{Hale:1994} that found relatively close binaries
(within 16 AU) to be spin-orbit aligned to within $\sim10\degrees$.
We argue that \citeauthor{Hale:1994}'s result could not have been obtained
if the derived inclinations were truly as uncertain as the formal
calculations imply.  With this point in mind, we find the case for
non-solar differential rotation in HD 81809A to be on firmer ground.
Such a scenario fits in with the recent proposition of
\citet{Brandenburg:2018} who hypothesized that the population of M67
dwarf stars with large Rossby number and relatively high activity may
in fact be examples of anti-solar differential rotation, as was
predicted by the models of \citet{Karak:2015}.  As we have discussed,
HD 81809A also likely has a large Rossby number as well as a high
activity compared to other similarly evolved stars.
}

The HD 81809 system has also been well studied in X-rays
\citep{Favata:2004,Favata:2008,Orlando:2017}.  \citet{Favata:2008}
assumed solar-like active regions and deduced that the primary
component must be emitting most of the flux, because under such an
assumption the secondary component would require an unphysical filling
factor larger than 1.  \citet{Radick:2018} did not find this argument
convincing, noting that younger, fast-rotating main-sequence stars
also have X-ray luminosities well exceeding the Sun's
\citep[e.g.][]{Wright:2011} with a similar surface area.  However, the
Ca HK and X-ray emission do vary in phase, and therefore it is
reasonable to assume that the $\sim$40 d rotation period measured in
the Ca timeseries also applies to the star responsible for the X-ray
cycle.  This is certainly \emph{not} the rotation period of a young
star, and therefore we believe the intuition of \citet{Favata:2004}
that the primary is the active component (reaffirmed in
\citet{Orlando:2017}) was ultimately correct.

\citet{Radick:2018} presented 19 years of photometric variability of HD
81809 in the Str\"{o}mgren $b$ (469 nm, FWHM 20 nm) and $y$ (548 nm,
FWHM 22 nm) bands.  It was found that photometric variability is
relatively large ($\sigma_{(b+y)/2} = 1.26$ mmag) and the
correlation between brightness and Ca HK activity is poor and
negative.  Compared to their best estimate for the Sun, HD 81809A is
2.6 times more variable in the $(b+y)/2$ bandpass.  The filling
factors used by \citet{Favata:2008} (later reconfirmed by
\citet{Orlando:2017}) to explain the X-ray luminosities were large;
60\% coverage by active regions and 4--40\% by brighter ``active
region cores'' from activity minimum to maximum.  This was done
assuming a radius of $R \sim 2 \, \Rnom$; rescaling to our value of \revone{2.42}
$\Rnom$ these are reduced somewhat to 38\% active regions and
\revone{2.7--27\%} by active region cores.  Based on our expectation of a
correspondence between coronal and photospheric features, we would
expect a strong cycle-scale variation in the photometry due to the
presence of such structures.  Therefore, as pointed out in
\citet{Radick:2018}, that we do not see a clear photometric cycle is
puzzling.  One possible explanation is that HD 81809A is on the
threshold between plage-dominated and spot-dominated photometric
variations, such that during times of high activity it is
spot-dominated while when activity is low it is plage dominated.
Observationally, this corresponds to the region where the change in
brightness with change in activity is near zero ($\Delta(b+y)/2 /
\Delta S) \sim 0$).  \citet{Radick:2018} finds this transition point
point at $\log(\RpHK) \sim -4.75$, though there is considerable
scatter about the zero-slope line for lower activity (see their Figure
15).  \citet{Shapiro:2014} used the SATIRE-S irradiance model to
predict the possibility of such transition-point stars, which depends
on their inclination, in the activity range $-4.9 \leq \log(\RpHK)
\leq -4.7$, which is near our estimated activity level for HD 81809A.

It is interesting to consider the magnetic evolution of a star at the
estimated mass of HD 81809A.  Inspection of Figure \ref{fig:tracks}
shows that HD 81809A was $\sim$1000 K warmer ($T_\eff \sim 6750$) when
it was on the main sequence.  This would put it at about the F3
spectral type with $(B-V) = 0.389$ \citep{Pecaut:2013}.  This is very
near the region of the H-R diagram where chromospheric emission (and
therefore magnetic activity) begins to be observed for main-sequence
stars, presumably coinciding with the onset of a thin outer convective
zone \citep{Bohm-Vitense:1980}.  Such a star is not expected to have
strong surface magnetic fields which cause stellar spin-down,
therefore it may spend the first 2--4 Gyr (depending on mass) at a
relatively fast rotation rate of $P_\rot \sim$ 3--5 days \citep[][see
  Figure 8]{VanSaders:2013}.  The rotation rate then rapidly decreases
as expansion takes place, and may reach $\sim$40 days in a fraction of
a Gyr \citep[][see Figures 8 and 9]{VanSaders:2013}\footnote{
  As discussed in \citet{VanSaders:2013}, subgiant rotation period can
  serve as a useful diagnostic.  From their radius vs. period diagram
  (Figure 9) we find that the HD 81809A should have a mass $M/\Mnom
  \sim 1.2$.  The \revone{slow} rotation then tightly constrains its age to
  $\sim 5$ Gyr, according to their age vs. rotation diagram (Figure
  8).  This radius-rotation based diagnostic is in good
  agreement with the MIST model mass and age discussed in Section
  \ref{sec:mist}\revone{, but is inconsistent with the estimate based
    on orbital solutions and parallax.}.
}.  Thus we now observe HD 81809A during a period of relatively rapid
magnetic evolution, as the rotation rate slows and the convection zone
expands.  The time period for which conditions suitable for a smooth
cycle such as HD 81809A now \revone{manifests} may be remarkably brief with
respect to its stellar lifetime.  Furthermore, the star was close to
magnetically \emph{inert} for its 2--4 Gyr main-sequence lifetime, and
likely became progressively more active as expansion produced a
thicker convective envelope.  Such magnetic evolution is in many ways
the opposite as that of the Sun, which began its main-sequence life
very magnetically active and then became progressively less so as it
spun down over $\sim$4 Gyr.  The distinctly different magnetic history
of HD 81809A and stars in a similar mass range may play a role in the
habitability and potential evolutionary history of their respective
exoplanets \citep[e.g.][]{Cuntz:2010}.

\vspace{1em}

I would like to thank Willie Soon and Jeffrey C. Hall for sharing the
$S$-index data for HD 81809.  Thanks also to Phil Judge, Travis
Metcalfe, and Scott McIntosh for the useful discussions on this work.
\revone{Thanks to the anonymous referee for the useful comments which
  improved the manuscript.}  This work was funded by the NCAR High
Altitude Observatory Newkirk Fellowship and the NCAR Advanced Study
Program Postdoctoral Fellowship.  The National Center for Atmospheric
Research is sponsored by the National Science Foundation.

\pagebreak

%\bibliography{bibdesk-master}
\bibliography{hd81809}

\begin{thebibliography}{}
\expandafter\ifx\csname natexlab\endcsname\relax\def\natexlab#1{#1}\fi

\bibitem[{{Allen}(1973)}]{Allen:1973}
{Allen}, C.~W. 1973, {Astrophysical Quantities}

\bibitem[{{Ammler-von Eiff} \& {Reiners}(2012)}]{AmmlerVonEiff:2012}
{Ammler-von Eiff}, M., \& {Reiners}, A. 2012, \aap, 542, A116

\bibitem[{{Baize}(1985)}]{Baize:1985}
{Baize}, P. 1985, \aaps, 60, 333

\bibitem[{{Baliunas} {et~al.}(1998){Baliunas}, {Donahue}, {Soon}, \&
  {Henry}}]{Baliunas:1998}
{Baliunas}, S.~L., {Donahue}, R.~A., {Soon}, W., \& {Henry}, G.~W. 1998, in
  Astronomical Society of the Pacific Conference Series, Vol. 154, The Tenth
  Cambridge Workshop on Cool Stars, Stellar Systems and the Sun, ed. R.~A.
  {Donahue} \& J.~A. {Bookbinder}, 153

\bibitem[{{Baliunas} {et~al.}(1995){Baliunas}, {Donahue}, {Soon}, {Horne},
  {Frazer}, {Woodard-Eklund}, {Bradford}, {Rao}, {Wilson}, {Zhang}, {Bennett},
  {Briggs}, {Carroll}, {Duncan}, {Figueroa}, {Lanning}, {Misch}, {Mueller},
  {Noyes}, {Poppe}, {Porter}, {Robinson}, {Russell}, {Shelton}, {Soyumer},
  {Vaughan}, \& {Whitney}}]{Baliunas:1995}
{Baliunas}, S.~L., {Donahue}, R.~A., {Soon}, W.~H., {et~al.} 1995, \apj, 438,
  269

\bibitem[{{Boeche} \& {Grebel}(2016)}]{Boeche:2016}
{Boeche}, C., \& {Grebel}, E.~K. 2016, \aap, 587, A2

\bibitem[{{Boehm-Vitense} \& {Dettmann}(1980)}]{Bohm-Vitense:1980}
{Boehm-Vitense}, E., \& {Dettmann}, T. 1980, \apj, 236, 560

\bibitem[{{B{\"o}hm-Vitense}(2007)}]{Bohm-Vitense:2007}
{B{\"o}hm-Vitense}, E. 2007, \apj, 657, 486

\bibitem[{{Brandenburg} \& {Giampapa}(2018)}]{Brandenburg:2018}
{Brandenburg}, A., \& {Giampapa}, M.~S. 2018, \apjl, 855, L22

\bibitem[{{Brandenburg} {et~al.}(2017){Brandenburg}, {Mathur}, \&
  {Metcalfe}}]{Brandenburg:2017}
{Brandenburg}, A., {Mathur}, S., \& {Metcalfe}, T.~S. 2017, \apj, 845, 79

\bibitem[{{Brandenburg} {et~al.}(1998){Brandenburg}, {Saar}, \&
  {Turpin}}]{Brandenburg:1998}
{Brandenburg}, A., {Saar}, S.~H., \& {Turpin}, C.~R. 1998, \apjl, 498, L51

\bibitem[{{Choi} {et~al.}(2016){Choi}, {Dotter}, {Conroy}, {Cantiello},
  {Paxton}, \& {Johnson}}]{Choi:2016}
{Choi}, J., {Dotter}, A., {Conroy}, C., {et~al.} 2016, \apj, 823, 102

\bibitem[{{Cuntz} {et~al.}(2010){Cuntz}, {Guinan}, \& {Kurucz}}]{Cuntz:2010}
{Cuntz}, M., {Guinan}, E.~F., \& {Kurucz}, R.~L. 2010, in IAU Symposium, Vol.
  264, Solar and Stellar Variability: Impact on Earth and Planets, ed. A.~G.
  {Kosovichev}, A.~H. {Andrei}, \& J.-P. {Rozelot}, 419--426

\bibitem[{{Donahue}(1993)}]{Donahue:1993:thesis}
{Donahue}, R.~A. 1993, PhD thesis, New Mexico State University, University
  Park.

\bibitem[{{Donahue} {et~al.}(1996){Donahue}, {Saar}, \&
  {Baliunas}}]{Donahue:1996}
{Donahue}, R.~A., {Saar}, S.~H., \& {Baliunas}, S.~L. 1996, \apj, 466, 384

\bibitem[{{Duquennoy} \& {Mayor}(1988)}]{Duquennoy:1988}
{Duquennoy}, A., \& {Mayor}, M. 1988, \aap, 195, 129

\bibitem[{{Egeland}(2017)}]{Egeland:2017:thesis}
{Egeland}, R. 2017, PhD thesis, Montana State University, Bozeman, Montana, USA

\bibitem[{{Egeland}(2018)}]{Egeland:2018b}
---. 2018, ArXiv e-prints, arXiv:1807.10870, submitted to \apj{}.

\bibitem[{{Egeland} {et~al.}(2017){Egeland}, {Soon}, {Baliunas}, {Hall},
  {Pevtsov}, \& {Bertello}}]{Egeland:2017}
{Egeland}, R., {Soon}, W., {Baliunas}, S., {et~al.} 2017, \apj, 835,
  doi:10.3847/1538-4357/835/1/25

\bibitem[{{ESA}(1997)}]{ESA:1997}
{ESA}, ed. 1997, ESA Special Publication, Vol. 1200, {The HIPPARCOS and TYCHO
  catalogues. Astrometric and photometric star catalogues derived from the ESA
  HIPPARCOS Space Astrometry Mission}

\bibitem[{{Fabricius} {et~al.}(2002){Fabricius}, {H{\o}g}, {Makarov}, {Mason},
  {Wycoff}, \& {Urban}}]{Fabricius:2002}
{Fabricius}, C., {H{\o}g}, E., {Makarov}, V.~V., {et~al.} 2002, \aap, 384, 180

\bibitem[{{Fabricius} \& {Makarov}(2000)}]{Fabricius:2000}
{Fabricius}, C., \& {Makarov}, V.~V. 2000, \aap, 356, 141

\bibitem[{{Favata} {et~al.}(2004){Favata}, {Micela}, {Baliunas}, {Schmitt},
  {G{\"u}del}, {Harnden}, {Sciortino}, \& {Stern}}]{Favata:2004}
{Favata}, F., {Micela}, G., {Baliunas}, S.~L., {et~al.} 2004, \aap, 418, L13

\bibitem[{{Favata} {et~al.}(2008){Favata}, {Micela}, {Orlando}, {Schmitt},
  {Sciortino}, \& {Hall}}]{Favata:2008}
{Favata}, F., {Micela}, G., {Orlando}, S., {et~al.} 2008, \aap, 490, 1121

\bibitem[{{Flower}(1996)}]{Flower:1996}
{Flower}, P.~J. 1996, \apj, 469, 355

\bibitem[{{Gaia Collaboration} {et~al.}(2016){Gaia Collaboration}, {Prusti},
  {de Bruijne}, {Brown}, {Vallenari}, {Babusiaux}, {Bailer-Jones}, {Bastian},
  {Biermann}, {Evans}, \& et~al.}]{Gaia:2016}
{Gaia Collaboration}, {Prusti}, T., {de Bruijne}, J.~H.~J., {et~al.} 2016,
  \aap, 595, A1

\bibitem[{{Gaia Collaboration} {et~al.}(2018){Gaia Collaboration}, {Eyer},
  {Rimoldini}, {Audard}, {Anderson}, {Nienartowicz}, {Glass}, {Marchal},
  {Grenon}, {Mowlavi}, {Holl}, {Clementini}, {Aerts}, {Mazeh}, {Evans},
  {Szabados}, \& {co-authors}}]{Gaia:2018:DR2}
{Gaia Collaboration}, {Eyer}, L., {Rimoldini}, L., {et~al.} 2018, ArXiv
  e-prints, arXiv:1804.09382

\bibitem[{{Gilliland}(1985)}]{Gilliland:1985}
{Gilliland}, R.~L. 1985, \apj, 299, 286

\bibitem[{{Gray} {et~al.}(2003){Gray}, {Corbally}, {Garrison}, {McFadden}, \&
  {Robinson}}]{Gray:2003}
{Gray}, R.~O., {Corbally}, C.~J., {Garrison}, R.~F., {McFadden}, M.~T., \&
  {Robinson}, P.~E. 2003, \aj, 126, 2048

\bibitem[{{Hale}(1994)}]{Hale:1994}
{Hale}, A. 1994, \aj, 107, 306

\bibitem[{{Hall} {et~al.}(2007){Hall}, {Lockwood}, \& {Skiff}}]{Hall:2007b}
{Hall}, J.~C., {Lockwood}, G.~W., \& {Skiff}, B.~A. 2007, \aj, 133, 862

\bibitem[{{Hartkopf} {et~al.}(2001){Hartkopf}, {Mason}, \&
  {Worley}}]{Hartkopf:2001}
{Hartkopf}, W.~I., {Mason}, B.~D., \& {Worley}, C.~E. 2001, \aj, 122, 3472

\bibitem[{{H{\o}g} {et~al.}(2000){H{\o}g}, {Fabricius}, {Makarov}, {Urban},
  {Corbin}, {Wycoff}, {Bastian}, {Schwekendiek}, \& {Wicenec}}]{Hog:2000}
{H{\o}g}, E., {Fabricius}, C., {Makarov}, V.~V., {et~al.} 2000, \aap, 355, L27

\bibitem[{{Holmberg} {et~al.}(2009){Holmberg}, {Nordstr{\"o}m}, \&
  {Andersen}}]{Holmberg:2009}
{Holmberg}, J., {Nordstr{\"o}m}, B., \& {Andersen}, J. 2009, \aap, 501, 941

\bibitem[{{Horne} \& {Baliunas}(1986)}]{Horne:1986}
{Horne}, J.~H., \& {Baliunas}, S.~L. 1986, \apj, 302, 757

\bibitem[{{Isaacson} \& {Fischer}(2010)}]{Isaacson:2010}
{Isaacson}, H., \& {Fischer}, D. 2010, \apj, 725, 875

\bibitem[{{Johnson} {et~al.}(1966){Johnson}, {Mitchell}, {Iriarte}, \&
  {Wisniewski}}]{Johnson:1966}
{Johnson}, H.~L., {Mitchell}, R.~I., {Iriarte}, B., \& {Wisniewski}, W.~Z.
  1966, Communications of the Lunar and Planetary Laboratory, 4, 99

\bibitem[{{Judge} {et~al.}(2003){Judge}, {Solomon}, \& {Ayres}}]{Judge:2003}
{Judge}, P.~G., {Solomon}, S.~C., \& {Ayres}, T.~R. 2003, \apj, 593, 534

\bibitem[{{Karak} {et~al.}(2015){Karak}, {K{\"a}pyl{\"a}}, {K{\"a}pyl{\"a}},
  {Brandenburg}, {Olspert}, \& {Pelt}}]{Karak:2015}
{Karak}, B.~B., {K{\"a}pyl{\"a}}, P.~J., {K{\"a}pyl{\"a}}, M.~J., {et~al.}
  2015, \aap, 576, A26

\bibitem[{{Linsky} \& {Avrett}(1970)}]{Linsky:1970}
{Linsky}, J.~L., \& {Avrett}, E.~H. 1970, \pasp, 82, 169

\bibitem[{{Livingston} {et~al.}(2007){Livingston}, {Wallace}, {White}, \&
  {Giampapa}}]{Livingston:2007}
{Livingston}, W., {Wallace}, L., {White}, O.~R., \& {Giampapa}, M.~S. 2007,
  \apj, 657, 1137

\bibitem[{{Lomb}(1976)}]{Lomb:1976}
{Lomb}, N.~R. 1976, \apss, 39, 447

\bibitem[{{Mamajek} \& {Hillenbrand}(2008)}]{Mamajek:2008}
{Mamajek}, E.~E., \& {Hillenbrand}, L.~A. 2008, \apj, 687, 1264

\bibitem[{{McAlister} {et~al.}(1990){McAlister}, {Hartkopf}, \&
  {Franz}}]{McAlister:1990}
{McAlister}, H., {Hartkopf}, W.~I., \& {Franz}, O.~G. 1990, \aj, 99, 965

\bibitem[{{Middelkoop}(1982)}]{Middelkoop:1982}
{Middelkoop}, F. 1982, \aap, 107, 31

\bibitem[{{Mishenina} {et~al.}(2013){Mishenina}, {Pignatari}, {Korotin},
  {Soubiran}, {Charbonnel}, {Thielemann}, {Gorbaneva}, \&
  {Basak}}]{Mishenina:2013}
{Mishenina}, T.~V., {Pignatari}, M., {Korotin}, S.~A., {et~al.} 2013, \aap,
  552, A128

\bibitem[{{Noyes} {et~al.}(1984){Noyes}, {Hartmann}, {Baliunas}, {Duncan}, \&
  {Vaughan}}]{Noyes:1984}
{Noyes}, R.~W., {Hartmann}, L.~W., {Baliunas}, S.~L., {Duncan}, D.~K., \&
  {Vaughan}, A.~H. 1984, \apj, 279, 763

\bibitem[{{Offner} {et~al.}(2016){Offner}, {Dunham}, {Lee}, {Arce}, \&
  {Fielding}}]{Offner:2016}
{Offner}, S.~S.~R., {Dunham}, M.~M., {Lee}, K.~I., {Arce}, H.~G., \&
  {Fielding}, D.~B. 2016, \apjl, 827, L11

\bibitem[{{Orlando} {et~al.}(2017){Orlando}, {Favata}, {Micela}, {Sciortino},
  {Maggio}, {Schmitt}, {Robrade}, \& {Mittag}}]{Orlando:2017}
{Orlando}, S., {Favata}, F., {Micela}, G., {et~al.} 2017, \aap, 605, A19

\bibitem[{{Pecaut} \& {Mamajek}(2013)}]{Pecaut:2013}
{Pecaut}, M.~J., \& {Mamajek}, E.~E. 2013, \apjs, 208, 9

\bibitem[{{Pourbaix}(2000)}]{Pourbaix:2000}
{Pourbaix}, D. 2000, \aaps, 145, 215

\bibitem[{{Pourbaix} {et~al.}(2004){Pourbaix}, {Tokovinin}, {Batten}, {Fekel},
  {Hartkopf}, {Levato}, {Morrell}, {Torres}, \& {Udry}}]{Pourbaix:2004}
{Pourbaix}, D., {Tokovinin}, A.~A., {Batten}, A.~H., {et~al.} 2004, \aap, 424,
  727

\bibitem[{{Pr{\v s}a} {et~al.}(2016){Pr{\v s}a}, {Harmanec}, {Torres},
  {Mamajek}, {Asplund}, {Capitaine}, {Christensen-Dalsgaard}, {Depagne},
  {Haberreiter}, {Hekker}, {Hilton}, {Kopp}, {Kostov}, {Kurtz}, {Laskar},
  {Mason}, {Milone}, {Montgomery}, {Richards}, {Schmutz}, {Schou}, \&
  {Stewart}}]{Prsa:2016}
{Pr{\v s}a}, A., {Harmanec}, P., {Torres}, G., {et~al.} 2016, \aj, 152, 41

\bibitem[{{Radick} {et~al.}(2018){Radick}, {Lockwood}, {Henry}, {Hall}, \&
  {Pevtsov}}]{Radick:2018}
{Radick}, R.~R., {Lockwood}, G.~W., {Henry}, G.~W., {Hall}, J.~C., \&
  {Pevtsov}, A.~A. 2018, \apj, 855, 75

\bibitem[{{Ram{\'{\i}}rez} {et~al.}(2012){Ram{\'{\i}}rez}, {Michel}, {Sefako},
  {Tucci Maia}, {Schuster}, {van Wyk}, {Mel{\'e}ndez}, {Casagrande}, \&
  {Castilho}}]{Ramirez:2012}
{Ram{\'{\i}}rez}, I., {Michel}, R., {Sefako}, R., {et~al.} 2012, \apj, 752, 5

\bibitem[{{Saar} \& {Brandenburg}(1999)}]{Saar:1999}
{Saar}, S.~H., \& {Brandenburg}, A. 1999, \apj, 524, 295

\bibitem[{{Shapiro} {et~al.}(2014){Shapiro}, {Solanki}, {Krivova}, {Schmutz},
  {Ball}, {Knaack}, {Rozanov}, \& {Unruh}}]{Shapiro:2014}
{Shapiro}, A.~I., {Solanki}, S.~K., {Krivova}, N.~A., {et~al.} 2014, \aap, 569,
  A38

\bibitem[{{S{\"o}derhjelm}(1999)}]{Soderhjelm:1999}
{S{\"o}derhjelm}, S. 1999, \aap, 341, 121

\bibitem[{{Takeda} \& {Kawanomoto}(2005)}]{Takeda:2005}
{Takeda}, Y., \& {Kawanomoto}, S. 2005, \pasj, 57, 45

\bibitem[{{Tokovinin} {et~al.}(2015){Tokovinin}, {Mason}, {Hartkopf}, {Mendez},
  \& {Horch}}]{Tokovinin:2015}
{Tokovinin}, A., {Mason}, B.~D., {Hartkopf}, W.~I., {Mendez}, R.~A., \&
  {Horch}, E.~P. 2015, \aj, 150, 50

\bibitem[{{Torres}(2010)}]{Torres:2010}
{Torres}, G. 2010, \aj, 140, 1158

\bibitem[{{van den Bos}(1938)}]{VanDenBos:1938}
{van den Bos}, W.~H. 1938, Circular of the Union Observatory Johannesburg, 100,
  481

\bibitem[{{van Saders} {et~al.}(2016){van Saders}, {Ceillier}, {Metcalfe},
  {Silva Aguirre}, {Pinsonneault}, {Garc{\'{\i}}a}, {Mathur}, \&
  {Davies}}]{VanSaders:2016}
{van Saders}, J.~L., {Ceillier}, T., {Metcalfe}, T.~S., {et~al.} 2016, \nat,
  529, 181

\bibitem[{{van Saders} \& {Pinsonneault}(2013)}]{VanSaders:2013}
{van Saders}, J.~L., \& {Pinsonneault}, M.~H. 2013, \apj, 776, 67

\bibitem[{{Vaughan} {et~al.}(1978){Vaughan}, {Preston}, \&
  {Wilson}}]{Vaughan:1978}
{Vaughan}, A.~H., {Preston}, G.~W., \& {Wilson}, O.~C. 1978, \pasp, 90, 267

\bibitem[{{Wilson}(1968)}]{Wilson:1968}
{Wilson}, O.~C. 1968, \apj, 153, 221

\bibitem[{{Wilson}(1978)}]{Wilson:1978}
---. 1978, \apj, 226, 379

\bibitem[{Wright(2004)}]{Wright:2004}
Wright, J.~T. 2004, The Astronomical Journal, 128, 1273

\bibitem[{{Wright} {et~al.}(2011){Wright}, {Drake}, {Mamajek}, \&
  {Henry}}]{Wright:2011}
{Wright}, N.~J., {Drake}, J.~J., {Mamajek}, E.~E., \& {Henry}, G.~W. 2011,
  \apj, 743, 48

\end{thebibliography}
\bibliographystyle{aasjournal}

\end{document}